\documentclass[%
reprint,
 amsmath,amssymb,
 aps,
prb,
floatfix,
]{revtex4-1}

\usepackage{graphicx}
\usepackage{dcolumn}
\usepackage{bm}
\usepackage{hyperref}
\usepackage[mathlines]{lineno}
\usepackage{xcolor}
\usepackage{hyperref}
\usepackage{cleveref}

\begin{document}

\title{Environment assisted and environment hampered efficiency at maximum power in a  molecular photo cell}

\author{Subhajit Sarkar}
\email{subhajit@post.bgu.ac.il}
 \affiliation{Department of Chemistry, Ben-Gurion University of the Negev, Beer-Sheva 8410501, Israel.}
\author{Yonatan Dubi}%
 \email{jdubi@bgu.ac.il}
\affiliation{%
Department of Chemistry, and the Ilse Katz Center for Nanoscale Science and Technology, Ben-Gurion University of the Negev, Beer-Sheva 8410501, Israel.
}


\date{\today}
\begin{abstract}

The molecular photo cell, i.e., a single molecule donor-acceptor complex, beside being technologically important, is a paradigmatic example of a many-body system operating in strong non-equilibrium. The quantum transport and the photo-voltaic energy conversion efficiency of the photocell, attached to two external leads, are investigated within the open quantum system approach by solving the Lindblad master equation. The interplay of the vibrational degrees of freedom corresponding to the molecules (via the electron-phonon interaction) and the environment (via dephasing) shows its signature in the efficiency at maximum power. We find vibration assisted electron transport in the medium to strong electron-phonon coupling regime when the system does not suffer dephasing. Exposure to dephasing hampers such a vibration assisted electron transport in a specific range of dephasing rate. 
\end{abstract}

\maketitle


\section{\label{sec:section11}Introduction}
Molecular junctions, i.e., single molecules or molecular layers placed between metal/semi-conductor electrodes is widely investigated due to it being a platform for studying fundamental aspects of non-equilibrium many body physics at the nano-meter scale, and its possible technological applications \cite{Aradhya2013, Galperin1056, Su2016, Evers2017,Thoss2018}. One such potential application (out of many) is the photo-voltaic (PV) cell, where the energy of incident photons is converted into electric current \cite{Gratzel2005, Deibel_2010,  Nicholson_2010}. In PV cells the interplay of the heat current originating from the temperature difference between the photon bath and the system (being at the ambient temperature), and the charge current originating from the bias voltage between the electrodes pushes the system to the strong non-equilibrium regime \cite{EDN2011, Ajisaka2015}. The operation of the molecular photo-cell has emerged as a rich theoretical problem, combining both fundamental understanding of quantum transport and  relevance to possible applications. \cite{Fruchtman2016, Arp2016, Killoran2015, Ernzerhof2016, Nemati-Aram2016, Nemati-Aram2017, Aram2017}

\paragraph*{}
The coupling of electronic degrees of freedom with the internal vibrations of the molecule may play a decisive role in charge transport though the molecular junction \cite{Ho2002, NitzanReview2001, Galperin_2007, Qin_Shen_Zhao_Yi_PhysRevA,Qin_Wang_Cui_Yi_PhysRevA,Qin_Shen_Yi_JChemPhys}. Analogous to the double slit experiment, in an electron only picture different pathways available for electron transfer in the molecular junctions interfere destructively to result in a small output current. Interactions with internal vibrations of molecules, available in the molecular junction, can quench the destructive interference between the electronic pathways thereby opening up more channels for transport, and lead to an enhanced current \cite{Ballmann2012, Hartle2011}.
\paragraph*{}
Coupling of electrons to the vibrations serves as an internal source of interruption of the phase coherent transport in molecular junctions. Apart from such a mechanism, phase coherent destructive interference of different electron transport channels can be destroyed by pure dephasing. This may result in enhanced transport current and efficiency, a phenomenon called  ``environment assisted quantum transport" (ENAQT) \cite{SowaRSC2017}. The phenomenon of ENAQT has been investigated quite extensively in the field of quantum biology in relation to the performance of photosynthetic systems \cite{Plenio_2008, Mohseni2008, Rebentrost_2009, Caruso_J_Chem_phys, Zerah-Harush2018}. Although the role of ENAQT in the performance of photo synthesis is much debated, its occurrence has recently been demonstrated in engineered quantum systems such as qubit, and optical cavity based networks \cite{Maier2019, Gorman2018, Viciani2015}. The conceptual similarity between natural photosynthetic systems and molecular photo-cells naturally raises the question - can the molecular environment (i.e. vibrations) assist the operation of a molecular photo-cell, and under which conditions?   

\paragraph*{} Here we investigate, by analysing the transport efficiency (at maximum power), the possibility of vibration assisted transport and ENAQT when the molecular junctions is used as a PV cell. It is shown that, within our generic model of a hetero-junction PV (HPV) cell the pure dephasing can indeed induce ENAQT. The situation of ENAQT in such a PV cell is quite different from that of a molecular junction placed in between leads in the sense that PV cell is continuously receiving photons from the solar radiation. Furthermore, in absence of dephasing vibration can also assist the electron transport through the junction.

It is worthwhile to point out that both vibration and dephasing are environmental effects on the HPV cell, the former being a localized environment and the latter characterizing an environment exhibiting a flat spectrum corresponding to the single dephasing rate. However, in the presence of both electron-vibration interactions and dephasing, a situation which to the best of our knowledge has not been investigated so far, the transport is actually hampered. Such a situation is quite realistic in view of the natural occurrence of vibration in molecular junctions and the noise induced dephasing inherent to a system placed in finite temperature.
\section{\label{sec:level2}Model}
\paragraph*{}
We consider an organic hetero-junction photo-voltaic cell (HPV cell) which is essentially a junction of coupled donor and acceptor molecules placed in-between two (left L, and right R) metallic leads. The minimal model for such an HPV cell consists of two ``effective" sites (the corresponding Hilbert space is denoted as $\mathcal{H}_{sites}$), the donor (D) and the acceptor (A) \cite{EDN2011, Ajisaka2015}. Both the donor and the acceptor are represented by two state systems with energy levels $(\epsilon_{D_1},\epsilon_{D_2})$ and $(\epsilon_{A_1},\epsilon_{A_2})$ corresponding to the (HOMO, LUMO) levels of them respectively. The most important energy scales in the system are HOMO-LUMO gap , $\Delta E = (\epsilon_{D_2}-\epsilon_{D_1})$, and the donor-acceptor band gap, $\Delta \epsilon = \epsilon_{D_2}-\epsilon_{A}$, where the acceptor HOMO ($\epsilon_{A_1}$) does not take part in the operation of the cell \cite{EDN2011}. The cell is placed under solar radiation which excites electrons from D-HOMO to D-LUMO (this process is termed as radiative transition). Moreover, the vibrational degrees of freedom, i.e., phonons of the donor molecule can cause non-radiative transitions between D-HOMO and D-LUMO.
\paragraph*{Molecular Hamiltonian:}
The quantum mechanical Hamiltonian for the above model of the HPV cell is written as $H = H_{e} + H_{e-ph}+
H_{ph}$. The electronic part of the total Hamiltonian is given by,
\begin{equation}\label{he}
    H_e = \sum_{i} \epsilon_i c_{i}^{\dagger} c_{i} - t (c_{D_2}^{\dagger} c_{A_2} + \text{h.c.}),
    \end{equation}
where $i = \{ D_1 , D_2 , A \}$, $c_i$ and $c_{i}^{\dagger}$ are the electron annihilation and creation operators respectively, and $t$ is the hopping amplitude between the D-LUMO and the A-LUMO.
\begin{figure}
\begin{center}
   \includegraphics[scale=0.25]{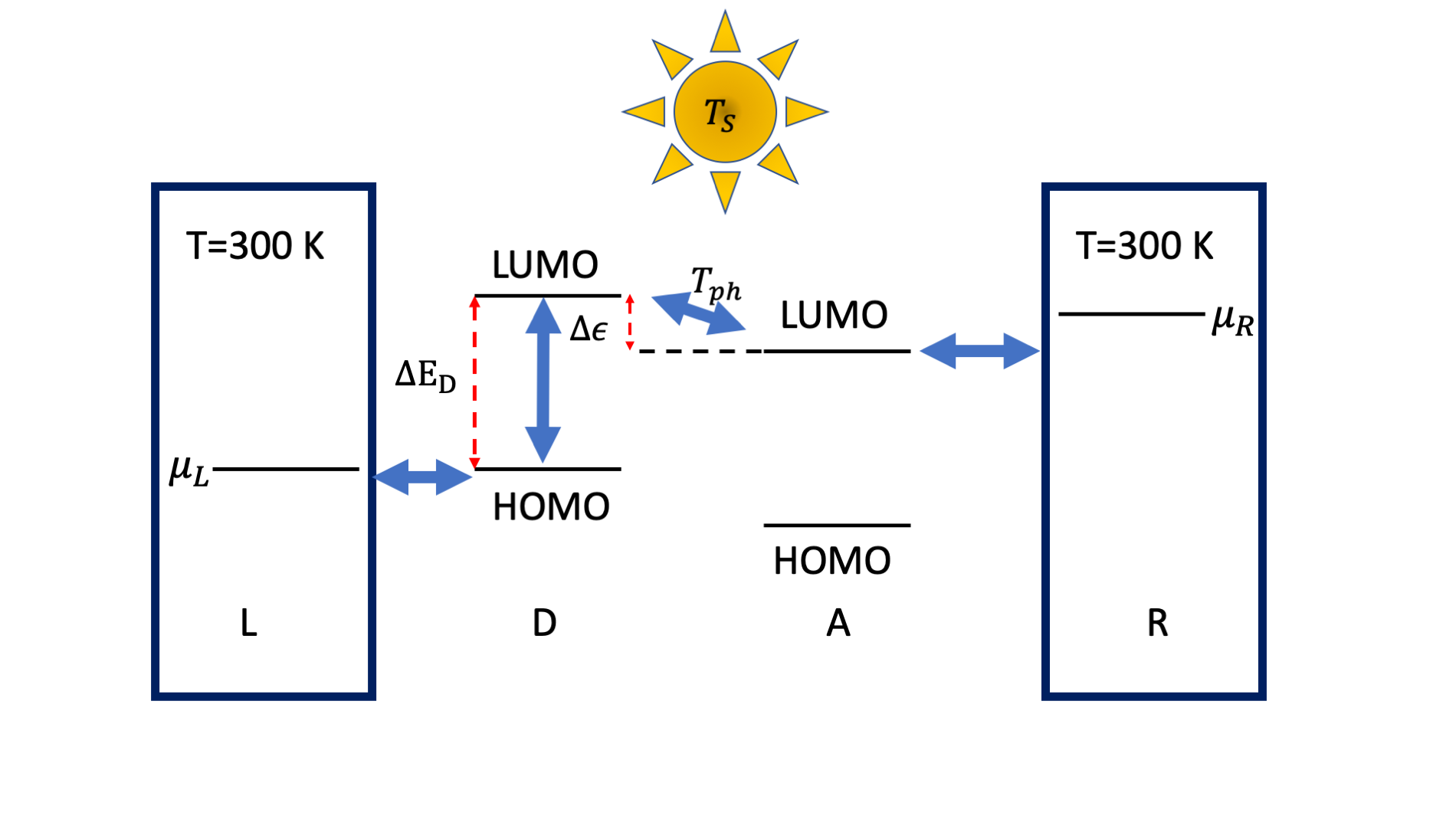} 
\end{center}
\caption{A minimal model for the molecular photo-voltaic cell incorporating the possible heating effect. Transfer of electron between the donor LUMO and the acceptor LUMO is controlled by two mechanisms, viz., electron hopping mechanism (characterized by an electron temperature $T_e$) and electron-phonon interaction ( characterized by temperature $T_{ph}$, as indicated in the figure). In our study we take $T_e = T_{ph} = 300$ K.}
\label{fig:model}
\end{figure}
\paragraph*{Electron-phonon interaction Hamiltonian:}
The interaction between the electrons of the D-A complex and the phonons are originated from the vibrational degrees of freedom of the parent molecules of the donor and the acceptor. The corresponding electron-phonon (e-ph) interaction Hamiltonian is given by, 

\begin{eqnarray}\label{heph}
      H_{e-ph} =\frac{\lambda_{e-ph}}{2} \left[ (b^{\dagger}+b) c_{A_2}^{\dagger} c_{D_2} + \text{h.c.}\right],
\end{eqnarray}
where $b \, (b^{\dagger})$ annihilates (creates) phonons, $\lambda_{e-ph}$ being e-ph coupling strength (a further simplification is obtained by taking the rotating wave approximation to this Hamiltonian). 
The transfer of electrons between the donor and the acceptor is therefore governed by two mechanisms, one being the electron hopping with strength `$t$', and the other being the non-radiative (e-ph) mechanism described by the above Hamiltonian which transfers electrons from donor to acceptor by creating or annihilating phonons. The Hamiltonian for the phonon bath is  $H_{ph} = \omega_{0} b^{\dagger} b $ with $\omega_{0} = 0.1 $ eV being the phonon energy, and the corresponding Hilbert space is denoted by $\mathcal{H}_{ph}$.
\paragraph*{}
The dynamics of the above mentioned system, i.e., the HPV cell with Hamiltonian $H$ is investigated within the Lindblad Master equation formalism, viz., $\dot{\rho} = -\frac{i}{\hbar} [H,\rho] + \sum_{j} (V_j \rho V_{j}^{\dagger} - \frac{1}{2} \{ V_{j}^  {\dagger} V_{j} , \rho \})$ where $\rho$ is the density matrix in the Hilbert space, $\mathcal{H}_{sites} \otimes \mathcal{H}_{ph}$ of the combined electron phonon system.  and $V_j$ are a set of Lindblad operators which encode the effects of the coupling to the environment \cite{Lindblad, GKS, Breuer, Archak_Dhar_Kulkarni_PhysRevA}. We model the photon and the phonon bath corresponding to the radiative and non-radiative transitions between the D-HOMO and the D-LUMO by suitable Lindblad $V$-operators. Such a model replaces the explicit appearance of phonon and photon baths considered in Ref. \cite{Ajisaka2015}, but mimics the relevant radiative and non-radiative transitions. The radiative part of the $V$-operators are given by,
\begin{eqnarray}
    V_{D_2 \rightarrow D_1}^{r} &=& \sqrt{\gamma_{pht} [n_{B}(\Delta E, T_s)+1]} c_{D_1}^{\dagger} c_{D_2} \nonumber \\
    V_{D_1 \rightarrow D_2}^{r} &=& \sqrt{\gamma_{pht} n_{B}(\Delta E, T_s)} c_{D_2}^{\dagger} c_{D_1},
\end{eqnarray}
where $n_{B}(\Delta E, T_s) = \frac{1}{\exp ( \frac{\Delta E}{k_B T_s} ) -1}$ is the Bose-Einstein distribution with Solar temperature $T_s = 5700$ K and the superscript `$r$' denotes the radiative part. $\gamma_{pht}$ is the rate of excitation or de-excitation of photons. We can write a similar pair of Lindblad $V$-operators for the non-radiative (denoted by superscript `$nr$') transition, viz., 
\begin{eqnarray}
    V_{D_2 \rightarrow D_1}^{nr} &=& \sqrt{\gamma_{phn} [n_{B}(\Delta E, T_{ph})+1]} c_{D_1}^{\dagger} c_{D_2} \nonumber \\
    V_{D_1 \rightarrow D_2}^{nr} &=& \sqrt{\gamma_{phn} n_{B}(\Delta E, T_{ph})} c_{D_2}^{\dagger} c_{D_1},
\end{eqnarray}
$\gamma_{phn}$ being the rate of excitation or de-excitation of phonons and $T_{ph}$ being the phonon temperature which we consider to be 300 K. In \textcolor{red}{\textit{Supplementary material}} 
we elaborate on and benchmark the above mentioned model of the photon and phonon baths against the model where the non-radiative (electron-phonon) and radiative (electron-photon interaction) processes between D-HOMO and D-LUMO explicitly appears in the Hamiltonian.

AN additional form of environmental influence is {\sl dephasing}, which is implemented via (zeno-type) measurement of local density. We consider here dephasing on the donor-LUMO and acceptor-HOMO states, which correspond to the Lindblad operators $V_{D_2} = \sqrt{\Gamma} c_{D_2}^{\dagger} c_{D_2} $ and $V_{A} = \sqrt{\Gamma}  c_{A}^{\dagger} c_{A}$, where $\Gamma$ is the dephasing rate. These represent the process of fast repetitive measurement of occupation at D-LUMO and A-LUMO respectively at the same rate \cite{SowaRSC2017, Plenio_2008, Mohseni2008, Rebentrost_2009, Caruso_J_Chem_phys, Zerah-Harush2018}. 

\paragraph*{}
We compute the efficiency at maximum power $\eta$, which is an important measure of the operational efficiency of the HPV cell \cite{EspositoPRL,Ajisaka2015,EDN2011}. This is defined as the ratio between the cell's maximal output power $P_{out}$, and the corresponding input power $P_{in}$ supplied by the photons, viz., $\eta = \frac{P_{out}[max]}{P_{in}[max]}$ \cite{VandenBroeckPRL, EspositoPRL}. The maximal output power $P_{out}[max]= J_{out}(V_{max}) V_{max}$, $J_{out}$ being the output current and the corresponding input power $P_{in} [max] = J_{S}(V_{max}) \Delta E$, where $J_S$ is the light induced current between the D-HOMO and the D-LUMO. See \textcolor{red}{\textit{supplementary material}} for the definitions of all the currents and powers.
\section{Results} 
\subsection{\textcolor{red}{\textit{ENAQT in absence of e-ph interaction}}}
We start by investigating the effects of dephasing on the efficiency at maximum power $\eta$ (henceforth the efficiency $\eta$) in absence of any e-ph interaction (\ref{heph}) whereby the system is described only by the Hamiltonian (\ref{he}). 
\begin{figure}
   \includegraphics[scale=0.5]{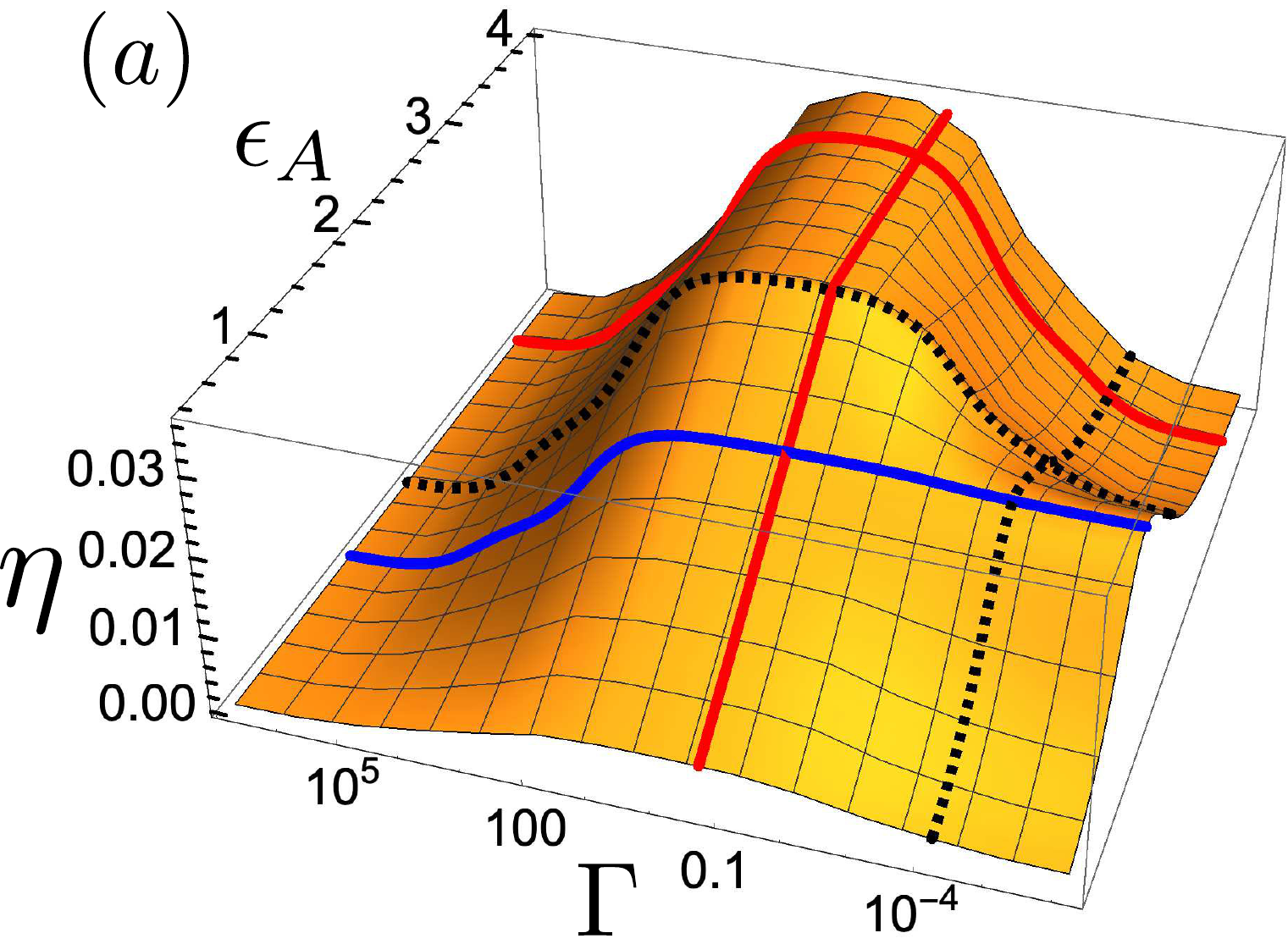} 
   \includegraphics[scale=0.8]{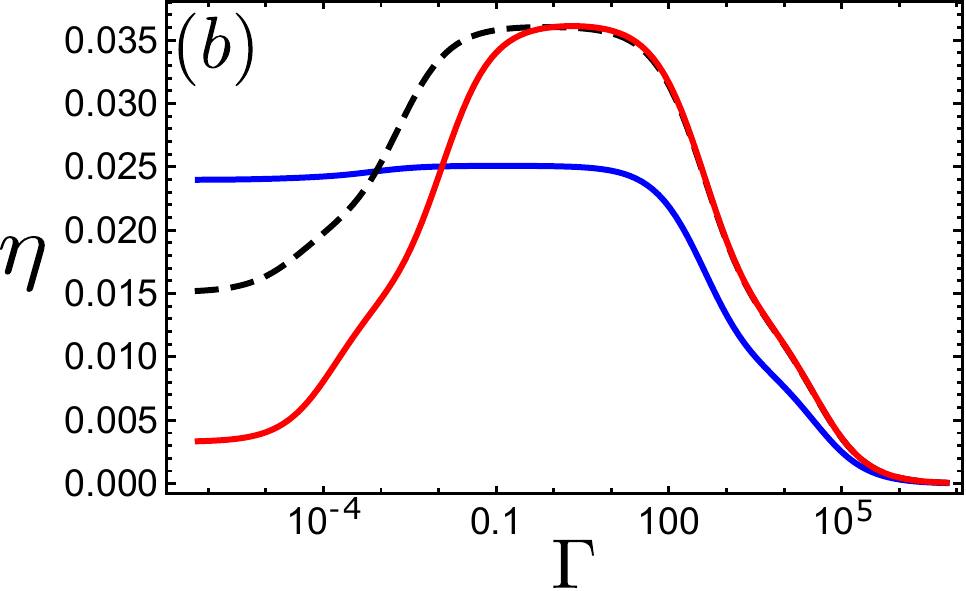} 
   \includegraphics[scale=0.8]{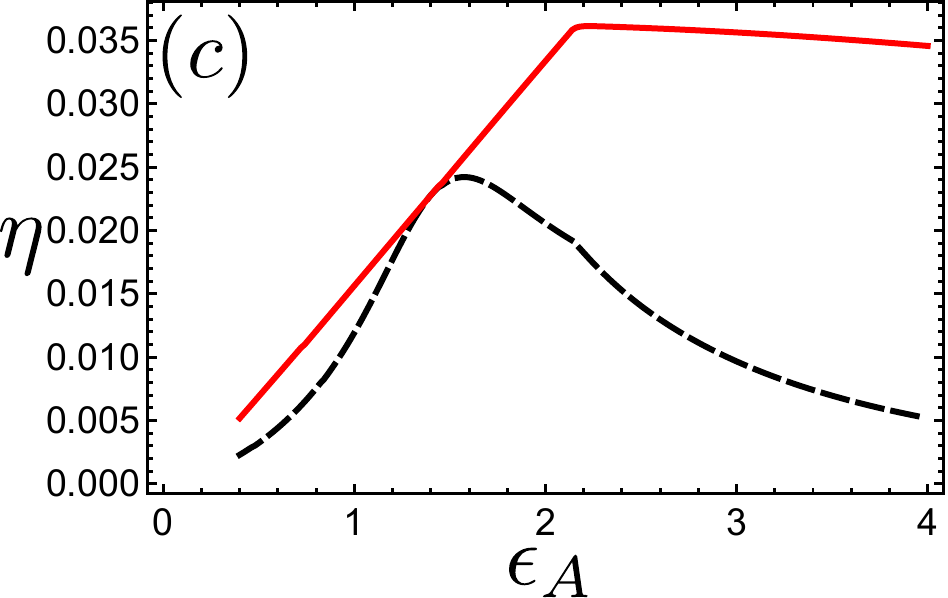} 
\caption{$(a)$ The density plot for the efficiency at maximum power $\eta$ as a function of acceptor energy level $\epsilon_A$ (in eV) and dephasing rate $\Gamma$ (in the units of eV). The color code represents the magnitude of $\eta$. $(b)$ Plot of $\eta$ as a function of $\Gamma$ for three different acceptor level positions, viz., $\epsilon_A = 1.53 $ eV (blue solid line), $2.15$ eV (black solid line) and $3.5$ eV (red solid line). These are indicated in $(a)$ by vertical solid lines. $(c)$ Plot of $\eta$ as a function of $\epsilon_{A}$ for dephasing rates $\Gamma = 10^{11} \hbar$ eV (black dashed line) and $10^{15} \hbar$ eV (red solid line). These are indicated in $(a)$ by horizontal dashed lines.}
\label{fig:eff0}
\end{figure}
In our calculation we set the orbital energies at $\epsilon_{D_1} = -0.1$ eV and $\epsilon_{D_2}= 1.4$ eV, the excitation rates $\gamma_{pht}=\gamma_{phn} = 10^{12} $ sec$^{-1}$, and the donor-acceptor electron hopping strength to be $t = 0.05$ eV. 

In Fig. \ref{fig:eff0} the efficiency at maximum power $\eta$ is plotted as a function of $\epsilon_{A}$ and dephasing rate $\Gamma$. We observe three distinct regime of dephasing rates along with the corresponding optimum values of $\epsilon_{A}$: (i) at small dephasing rate $\Gamma \lessapprox 10^{11} \hbar$ (in eV) the system is close to a quantum one with optimum $\epsilon_{A} = 1.53 $ eV. In this regime the dynamics of the system is dominated by the electron hopping strength $t$. 

(ii) When dephasing rate increases to a critical regime where $\Gamma$ becomes comparable to $\epsilon_{A}$, the donor-acceptor electron hopping mechanism combines with the dephasing to increase the efficiency of electron transfer. This regime corresponds to the ENAQT (enviornment assisted quantum transport) regime with the optimum $\epsilon_{A} = 2.15$ eV \cite{Rebentrost_2009, Caruso_J_Chem_phys}. 

(iii) A third regime corresponding to $\Gamma \gtrapprox (10^{20} \hbar)$ eV where the system become classical and all the electron transfers are reduced by the dephasing process. In this case the efficiency drops substantially (becoming roughly inversely proportional to the dephasing rate  \cite{dziarmaga2012non}). It is worthwhile to point out that the width of the ENAQT regime is controlled by the rate of injection of electrons at the D-LUMO level, which in our system is determined by the resultant of radiative ($\gamma_{pht}$) and non-radiative ($\gamma_{phn}$) transfer rates \cite{zerahharush2019environmentassisted}.

\subsection{ \textcolor{red}{\textit{Efficiency at the maximum power in presence of e-ph interaction: phonon assisted transport}}} 
\begin{figure}
\begin{center}
   \includegraphics[scale=0.30]{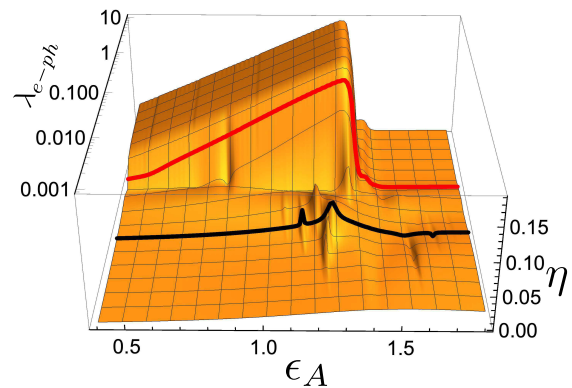} 
\end{center}
\caption{Efficiency at maximum power $\eta$ as a function of acceptor energy level $\epsilon_A$ (in eV) and e-ph coupling strength $\lambda_{eph}$ (in eV). The color code represents the magnitude of $\eta$. $(b)$ The black solid line and the red solid line correspond to the plot of $\eta$ as a function of $\epsilon_{A}$ for $\lambda_{e-ph} = 0.032 $ eV and $0.4$ eV respectively.}
\label{fig:eff1}
\end{figure}

Having established the presence of ENAQT in our system,  we now investigate the effect of e-ph interaction (corresponding to the Hamiltonian (\ref{heph})) on the efficiency at maximum power $\eta$, but without exposing the system to environment (i.e., without taking into account the dephasing effect). In Fig. \ref{fig:eff1} the efficiency at maximum power $\eta$ is plotted as a function of $\epsilon_{A}$ and $\lambda_{e-ph}$. From Fig. \ref{fig:eff1} it can be seen that for smaller values of $\lambda_{e-ph} \lesssim 0.01$ eV the efficiency $\eta$ peaks at $\epsilon_{A} = 1.25$ eV. When $\lambda_{e-ph}$ increases more than 0.01 eV, $\eta$ develops a two peak structure with maximum of $\eta$ for $\lambda_{e-ph} = 0.032$ eV.

\begin{figure}
\begin{center}
    \includegraphics[scale=0.45]{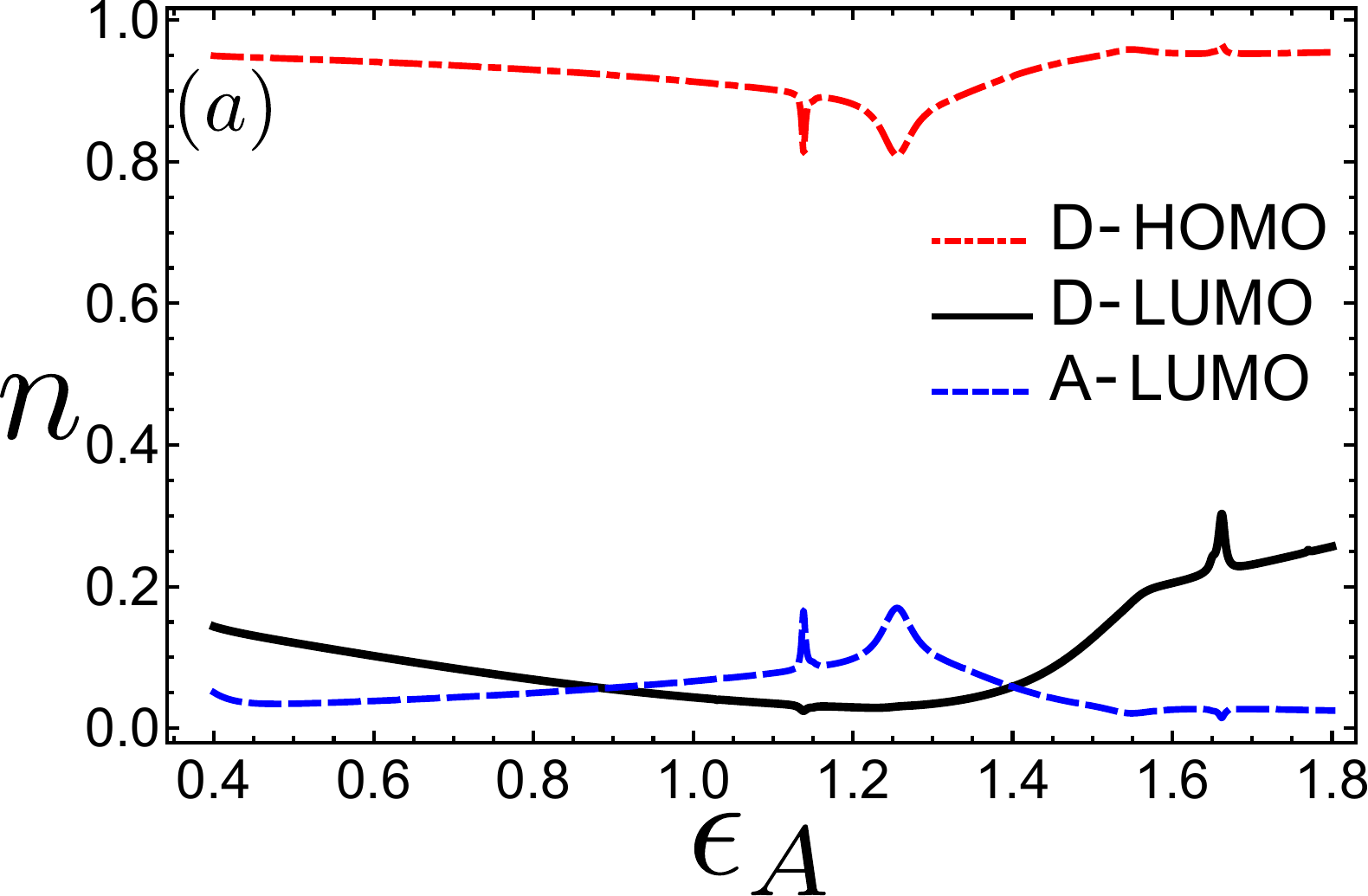}
    \includegraphics[scale=0.45]{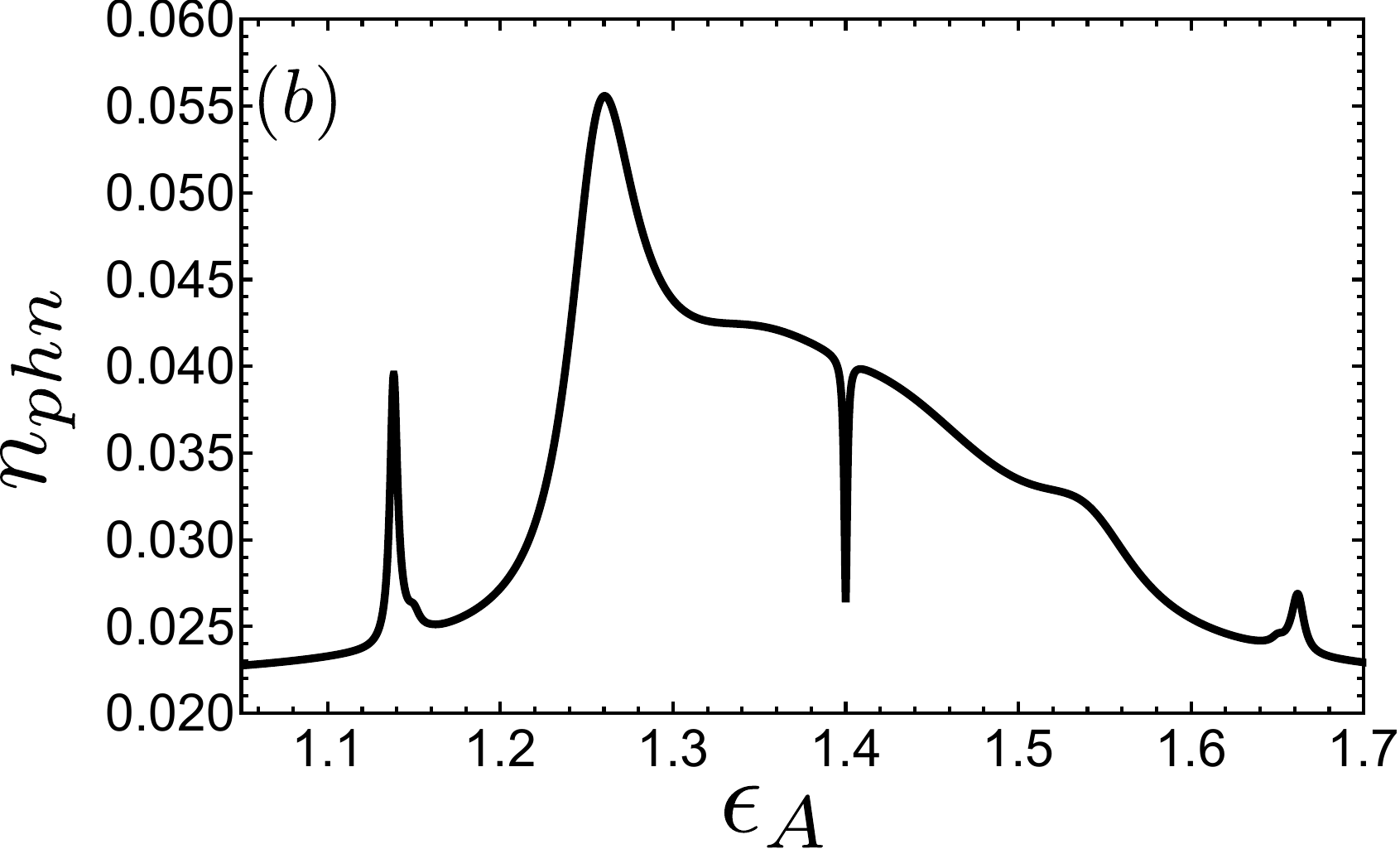}
\end{center}
\caption{$(a)$ Occupation numbers corresponding to D-HOMO, D-LUMO, and A-LUMO and $(b)$ plot of the phonon density as a function of $\epsilon_{A}$ for hopping strength $t = 0.05$ eV and e-ph coupling strength $\lambda_{e-ph} = 0.32 $ eV. These correspond to the black solid line of Fig. \ref{fig:eff1}.}
\label{fig:eff2}
\end{figure}

Such a two peak structure of the efficiency at maximum power $\eta$ as a function of the acceptor level $\epsilon_{A}$ can be understood by considering the dependence of the occupation on the A-LUMO positions. The electron occupations ($n_{D-HOMO}$, $n_{D-LUMO}$, and $n_{A-LUMO}$), and phonon occupation \textcolor{red}{\textit{(see supplementary material for the definitions of the electron and phonon occupations)}} are plotted as functions of $\epsilon_{A}$ in Fig. \ref{fig:eff2} $(a)$ and $(b)$ respectively, for $t=0.05$ eV and $\lambda_{e-ph} = 0.032$ eV. Careful observation of Fig. \ref{fig:eff2} $(a)$ and $(b)$ in comparison with the black solid curve corresponding to Fig. \ref{fig:eff1} reveals that peaks in $\eta$ appear at those values of $\epsilon_{A}$ for which $n_{A-LUMO}$ gets its peaks and both the $n_{D-HOMO}$ and $n_{D-LUMO}$ get its local dips. 

This correspondence between peak positions of efficiency and populations indicates that at these positions of A-LUMO, transfer of electrons to A-LUMO is maximized, leading to an increased efficiency. These are the electrons which get excited from the D-HOMO due to solar radiation (duly counterbalanced by the non-radiative re-combinations), and at the above mentioned values of $\epsilon_{A}$ (corresponding to the two peaks in $\eta$) the rate of excitation becomes maximal leading to dips in $n_{D-HOMO}$. Subsequently, the excited electrons are transferred to A-LUMO via D-LUMO leading to almost feature-less $n_{D-LUMO}$ near the two peaks of $\eta$. Furthermore, Fig.\ref{fig:eff2} $(b)$ indicates that phonon occupation also gets its local peaks at the same values of $\epsilon_{A}$. This feature leads us to infer that the transport here is phonon-assisted.

\begin{figure}
\begin{center}
   \includegraphics[scale=0.32]{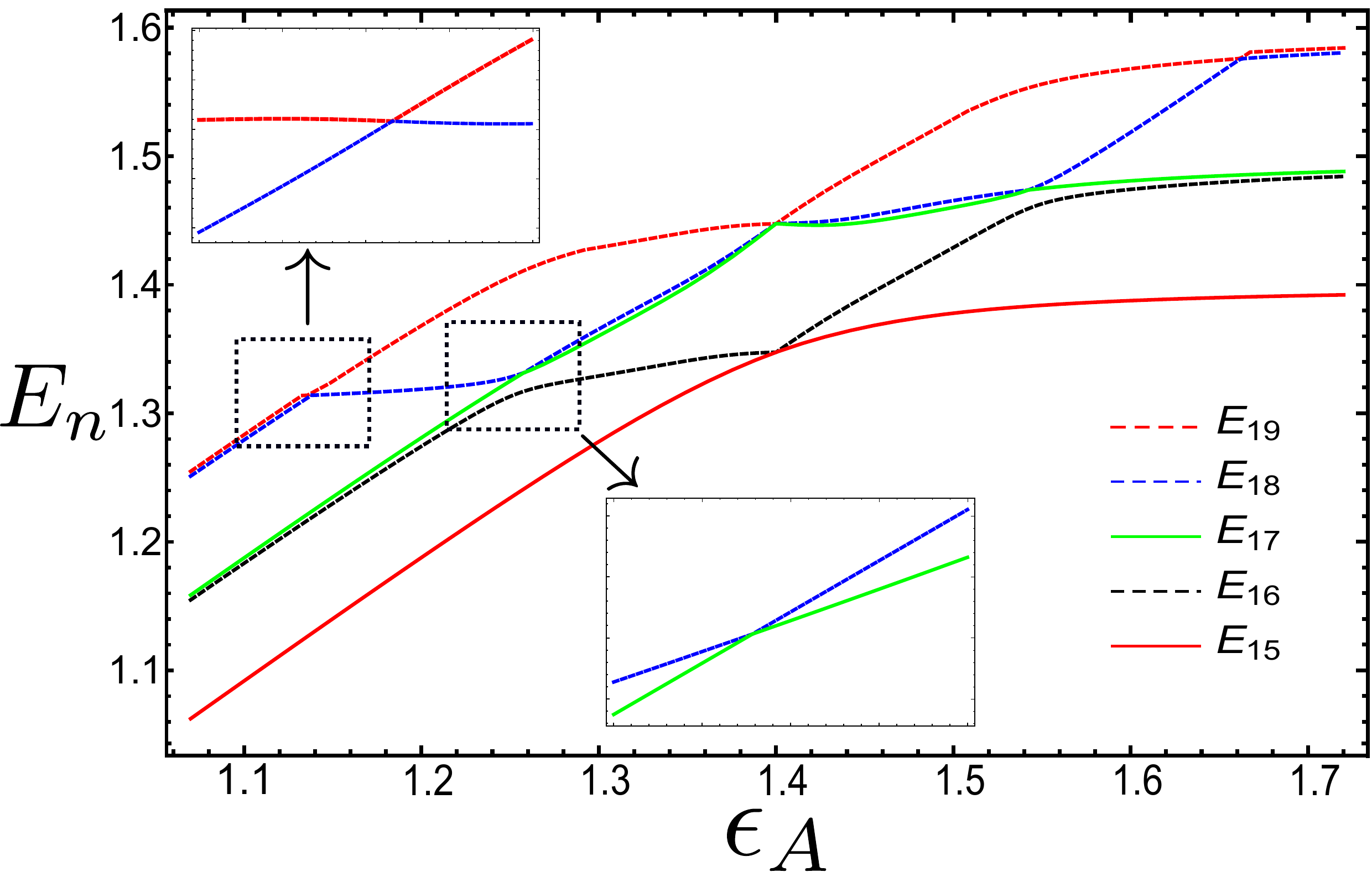} 
\end{center}
\caption{Plot of the real part of the eigenvalues $E_n$ of the $n$'th eigenstate of the non-Hermitian Hamiltonian $H_{eff}$ mentioned above. All the level crossing appearing in the plot occur at the values of $\epsilon_{A}$ for which $n_{ph}$ of Fig. \ref{fig:eff2} (b) shows either a local peak or a local dip. At $\epsilon_{A} = 1.14 $ eV the level crossing occurs between $E_{18}$ and $E_{19}$, and at $\epsilon_{A} = 1.25 $ eV the level crossing occurs between $E_{17}$ and $E_{18}$ and these are the crossing corresponding to the two local peaks in Fig. \ref{fig:eff1}($b$).}
\label{fig:eigen}
\end{figure}

\paragraph*{}
It is worthwhile to point out that $n_{ph}$ exhibits a resonant dip at $\epsilon_{A} = \epsilon_{D_2}$ $(= 1.4$ eV) and correspondingly $n_{D-LUMO}$ becomes more than $n_{A-LUMO}$ (at $\epsilon_{A} = \epsilon_{D_2}$ they becoming equal). On the contrary, there exists another such crossing at $\epsilon_{A} = 0.9$ eV where $n_{A-LUMO}$ becomes more than $n_{D-LUMO}$ where the $n_{ph}$ remains featureless because of the fact that at this position of the acceptor level no resonance condition (such as $\epsilon_{A} = \epsilon_{D_2}$) gets satisfied. 
\paragraph*{}
To substantiate our interpretation of the phonon assisted transport, we argue that such a two peak structure (in efficiency vs acceptor energy) is a resonant effect inherited from the level crossing of the eigenvalues (real part) of the effective non-Hermitian Hamiltonian $H_{\text{eff}} = H - \frac{i}{2} \sum_{j} \gamma_{j} V_{j}^{\dagger} V_{j}$, where $\sqrt{\gamma_{j}}$'s are the rates associated to the Lindblad operators $V_{j}$'s. The real part of a few eigenvalues of $H_{\text{eff}}$ are plotted in Fig. \ref{fig:eigen} as a function of acceptor level's energy $\epsilon_{A}$, the imaginary part being the life-time of the system to remain in the corresponding eigenstate. Only  eigenstates for which the eigenvalues are within the range of our interest, i.e., within he range of values of $\epsilon_{A}$ where $\eta$ shows the two peak structure, are shown.

As seen from Fig. \ref{fig:eigen}, the system exhibits several level crossings as the acceptor energy (hence acceptor position) is increased gradually to pass across the D-LUMO ($\epsilon_{D_2}$) level. Comparing Figs. \ref{fig:eff2} (a) and (b) with Fig. \ref{fig:eigen} one can clearly see the $\epsilon_{A} = 1.14$ and $1.25$ eVs level crossings, associated with the two peak structure of $\eta$. 
\begin{figure*}
\begin{center}
   \includegraphics[scale=0.35]{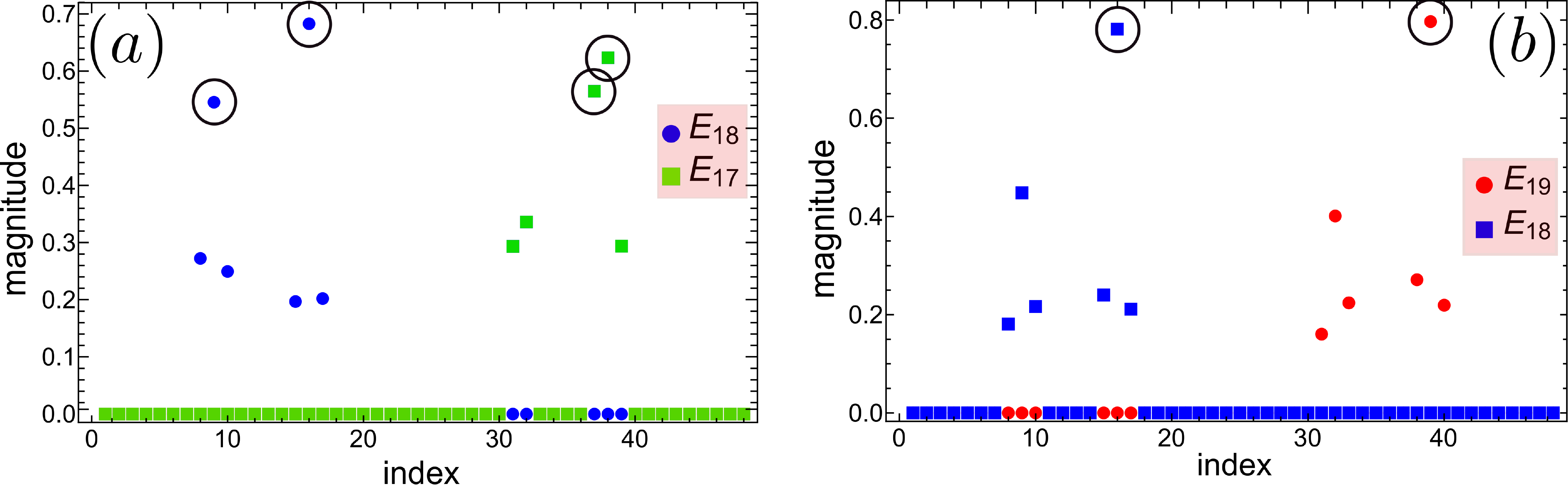} 
\end{center}
\caption{Scatter plot for the eigenvectors corresponding to the eigenvalues which cross each other ($a$) at $\epsilon_{A} = 1.25$ eV,  and ($b$) at $\epsilon_{A} = 1.14$ eV . The encircled ones represent the dominant system configuration contributing to the eigen-state.}
\label{fig:config}
\end{figure*}

In order to analyze the nature of transport near these resonances, in Fig. \ref{fig:config} the eigenvectors corresponding to the levels participating in the crossings are plotted as scatter plots;  the index corresponding to the encircled dots represent the diagonal elements of the Hamiltonian $H$ which in turn represent the most dominant system configuration contributing to the eigenstate. Considering the fact that the eigenstates can be written as $| \psi_{n} \rangle = \sum_{j=1}^{N} a_{n, j} | e_{j} \rangle$ where $N=2^3 \times N_{ph}$ is the dimension of the Hilbert space for truncated Bosonic space containing $N_{ph}$ number of phonons, and $|e_{j} \rangle$ are the unit vectors in the Hilbert space $\mathcal{H}_{sites} \otimes \mathcal{H}_{ph}$, such a configuration can be thought of as a semi-classical configuration of the system. This is in the sense that we are taking only a few $a_{j}$s and the corresponding $|e_{j}\rangle$s as the system configuration.

At $\epsilon_{A} = 1.14$ eV for the level crossing from $E_{18}$ to $E_{19}$ the encircled dots represent a transition from the system configuration $\epsilon_{D_1} + \epsilon_{A} +3\omega_{0}$ to $\epsilon_{A} + 2\omega_{0}$ indicating a transfer of one electron from donor to the right lead via the acceptor and a subsequent emission of one phonon of energy $\omega_0$. It is worthwhile to mention that within our model, the participation of the high energy phonons dominates the transport. At $\epsilon_{A} = 1.25$ eV for the level crossing from $E_{17}$ to $E_{18}$ the corresponding encircled dots represent a transition from ($\epsilon_{D_1} + \epsilon_{A} +3\omega_{0}$, $\epsilon_{D_1} + \epsilon_{D_2} +2\omega_{0}$) to ($\epsilon_{A} + \omega_{0}$, $\epsilon_{A}$) 
and subsequent emission of two phonons of energy $\omega_0$. Therefore at the first peak in $\eta$ the transport involves a one phonon process and while at the second peak it involves two phonon processes. Furthermore, at both the values of $\epsilon_{A}$, phonon emissions are involved in the electrons transfer processes and such emissions indeed increase the phonon occupation as seen from corresponding peaks in $n_{phn}$ in Fig. \ref{fig:eff2} $(b)$.

\paragraph*{}

Going back to Fig.~\ref{fig:eff1};  for large enough e-ph coupling strength corresponding to $\lambda_{eph} \gtrapprox 0.3$ eV (henceforth considered as strong e-ph coupling regime), the maximum of the efficiency $\eta$ is found at the optimum $\epsilon_{A} = 1.32$ eV. The maximum value of $\eta$ is around five fold more than what can be obtained in an electron-only mechanism of charge transport in HPV cells even in the ENAQT regime. This is due to the (rather well-known) fact that e-ph coupling opens up more channels for electron to get transferred from D-LUMO to A-LUMO, and in the strong e-ph coupling regimes these channels dominantly participate in transferring electrons leading to the enhanced efficiency. Furthermore, the optimum value of $\epsilon_{A}$, mentioned above, is weakly dependent of the e-ph coupling strength in the strong e-ph coupling regime. The intuitive way to understand why the maximum of $\eta$ occurs at this particular value of $\epsilon_A$ is as follows. From the red solid line corresponding to $\lambda_{e-ph} = 0.4$ eV in Fig. \ref{fig:eff1} $(b)$ we see that $\eta$ keeps increasing as the absolute position of the acceptor level approaches to the absolute position of the D-LUMO (in other words $\Delta \epsilon$ decreases). This happens because, with the A-LUMO approaching the D-LUMO the overlap between the wave-functions of these two levels increases, and consequently the phonon mediated transfer of electrons from D-LUMO to A-LUMO also increases. This results in an increased efficiency. As long as $\epsilon_A < \epsilon_{D_2}$ the electrons from D-LUMO gets transferred to A-LUMO by creating a phonon mode. When the situation $\epsilon_A > \epsilon_{D_2}$ appears the phonon mediated mechanism ceases to occur and the efficiency drops.
\subsection{\textcolor{red}{\textit{Environment hampered transport in the presence of e-ph interaction}}}
We now expose our combined electron phonon system to the environment via applying dephasing on both D-LUMO and A-LUMO at the same rate. We consider the e-ph coupling strength to be $\lambda_{e-ph} = 0.4$ eV, a case in which the efficiency at maximum power $\eta$ exhibits a maximum at $\epsilon_{A} = 1.32$ eV, as can be seen from Fig. \ref{fig:eff1} ($b$).

\begin{figure}
\begin{center}
   \includegraphics[scale=0.4]{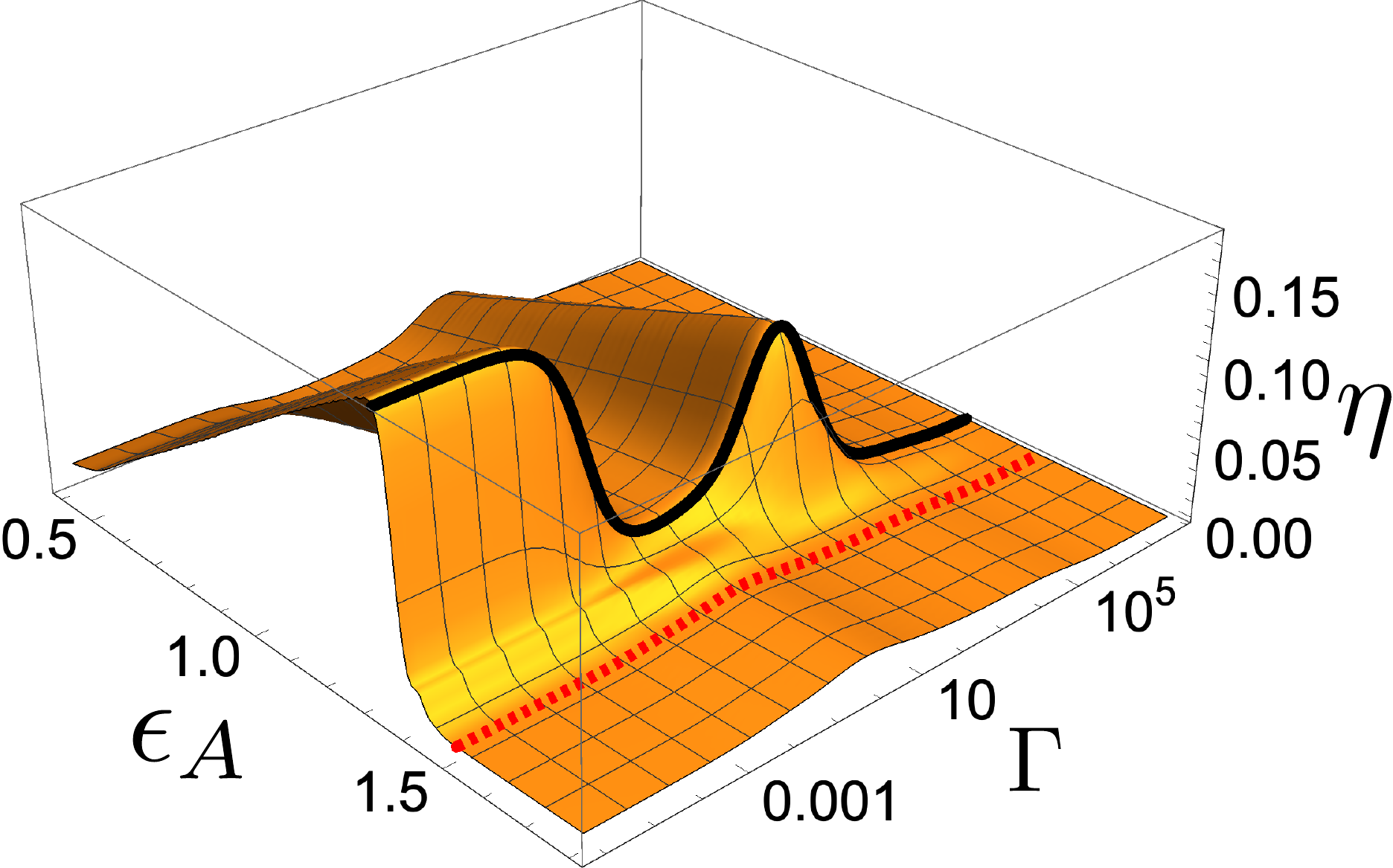} 
\end{center}
\caption{ The plot of efficiency at maximum power $\eta$ as a function of acceptor energy level $\epsilon_A$ (in eV) and dephasing rate $\Gamma$ (in eV) for $\lambda_{e-ph} = 0.4 $ eV. The black solid line and the red dashed line correspond to plot of $\eta$ as a function of  $\Gamma$ for  $\epsilon_{A} = 1.32$ eV and $1.45$ eV respectively. The local minimum appearing in the black solid line is at $\Gamma = 0.51$ eV.}
\label{fig:eff3}
\end{figure}

\begin{figure}[h]
\begin{center}
  \includegraphics[scale=0.35]{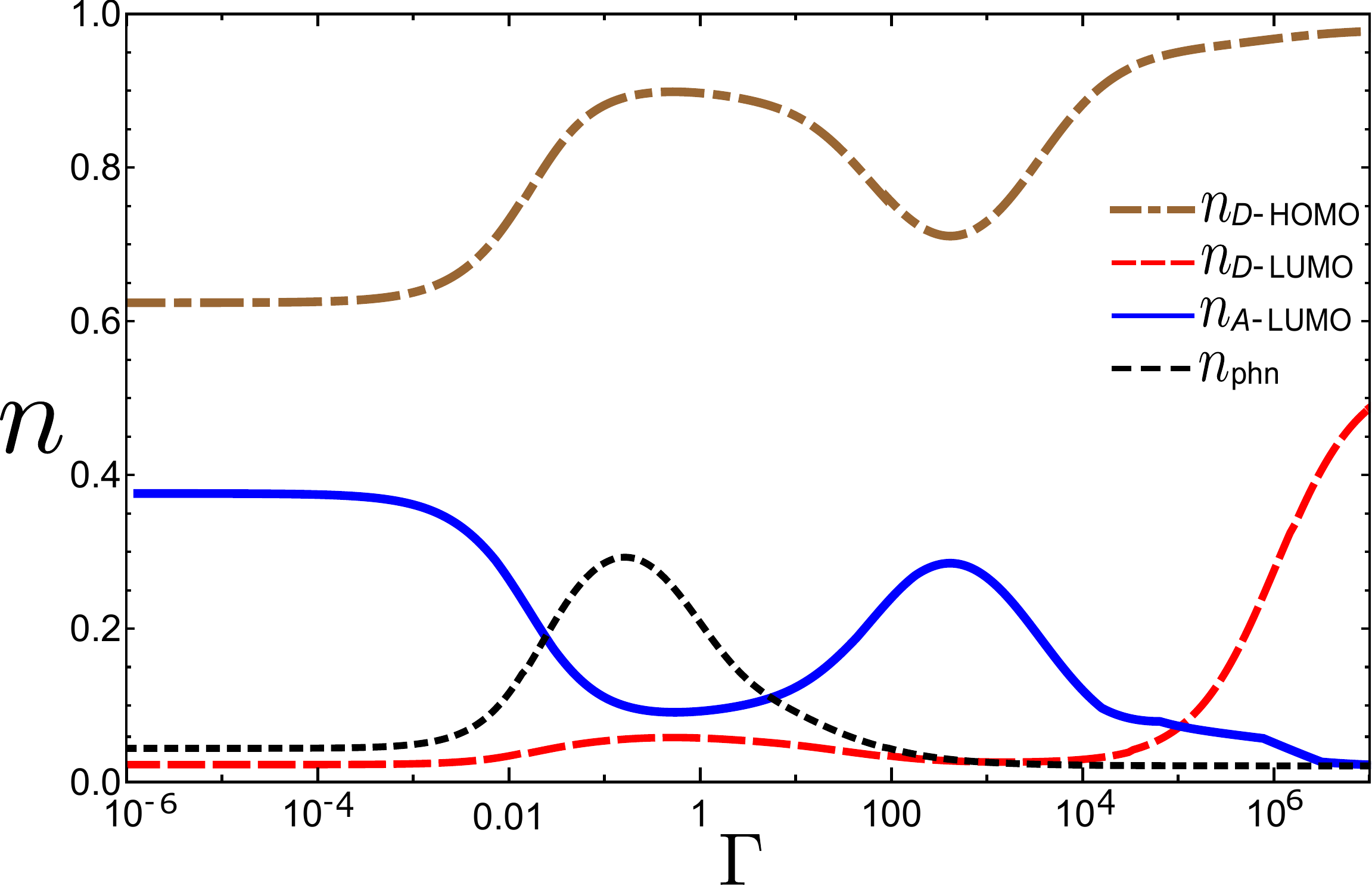} 
\end{center}
\caption{The plot of the electron occupations, $n_{D-HOMO}$ (brown dot-dashed line), $n_{D-LUMO}$ (red long-dashed line), and $n_{A-LUMO}$(blue solid line), and phonon occupation $n_{phn}$ (black short-dashed line) at maximum power as functions of dephasing rate $\Gamma$ (in eV) for bias voltages for which the output power is maximum.}
\label{fig:occn_dephase}
\end{figure}

In Fig. \ref{fig:eff3}, $\eta$ is plotted as a function of the acceptor energy $\epsilon_{A}$ and the dephasing strength $\Gamma$. For $\epsilon_{A} = 1.32$ eV (for which the transport is dominated by the e-ph interaction) we can clearly see an environment hampered quantum transport. This can be identified from the local dip in the $\eta$ in the regime of the dephasing rate which is otherwise the regime of ENAQT when the mechanism of transport was electron hopping only. The red dashed curve corresponding to Fig. \ref{fig:eff3} provides $\eta$ as a function of dephasing rate $\Gamma$ for $\epsilon_{A} = 1.45$ eV where the transport is due to electron hopping mechanism as the e-ph mechanism ceases to exist. A subsequent peak in the higher dephasing rate can bee seen however, at this peak $\eta$ remains less than it's values corresponding to the low dephasing regime corresponding to $\Gamma \lesssim 10^{12}\hbar$. This points to the fact that the ENAQT is never achieved when both the e-ph mechanism of transport and the dephasing are present together.
\paragraph*{}
 This environment hampered transport can be understood from the behavior of both the electron and the phonon occupations as a function of $\Gamma$. In Fig. \ref{fig:occn_dephase}, the occupation of electrons in D-HOMO, D-LUMO, and A-LUMO and phonon occupation $n_{phn}$ at maximum power are plotted as a function of dephasing rate $\Gamma$. First of all, comparing Figs. \ref{fig:eff3} and \ref{fig:occn_dephase} it can be seen that the behaviour of $\eta$ as a function of $\Gamma$ is same as that of the $n_{A-LUMO}$, indicating that $\eta$ is directly proportional to the A-LUMO occupation at maximum power.
 
 Moreover, three distinct regimes of transport are observed: (i) At small dephasing rate $\Gamma \lesssim 10^{12} \hbar$
electron density in A-LUMO is more than D-LUMO occupation and phonon occupation, although the dephasing rate $\Gamma$ is the same for both D-LUMO and A-LUMO. Therefore, more electrons are available at A-LUMO to be absorbed in the right lead which results in an increased $\eta$. In this regime electrons are transferred from D-HOMO to A-LUMO via D-LUMO by creating phonon modes, virtually spending no time in the D-LUMO.

(ii) For $\Gamma \gtrsim 10^{12} \hbar$ the phonon occupation starts increasing and the A-LUMO occupation starts decreasing which indicates with the increased dephasing the phonon bath gets populated by de-populating the A-LUMO. Dephasing, being a process mimicking the physical process of repeated measurements, tends to populate equally both the D-LUMO and A-LUMO by de-localizing electrons in the respective levels. In this process it essentially blocks most of those transport channels, which were otherwise available in the low dephasing regime due to electron phonon coupling, thereby reducing the phonon assisted transport of electrons from donor to acceptor resulting in a reduced $n_{A-LUMO}$. Therefore, the transfer of electrons from D-LUMO to A-LUMO is hampered by dephasing (therefore, by the environment). In this situation all the absorbed solar energy gets utilized to increase the $n_{phn}$, as can be seen from Fig. \ref{fig:occn_dephase}. 

Furthermore, in this regime corresponding to $10^{12} \hbar < \Gamma < 10^{18} \hbar$, the dephasing also hampers the effective injection of electrons in the D-LUMO level, as can be seen from the increased electron occupation in D-HOMO in this regime. When the dephasing rate exceed that of the transfer of electrons from D-HOMO to D-LUMO, it traps electrons in D-LUMO at a rate faster than the rate of excitation of electrons to D-LUMO leading to increased D-HOMO occupation. At resonant dephasing rate, corresponding to $\Gamma_{R} = 0.51$eV, electron density in A-LUMO is minimized locally in order to make
 occupation in A-LUMO and D-LUMO nearly equal (in this situation dephasing tends to populate both D-LUMO and A-LUMO equally \citep{Rebentrost_2009}). However, this is achieved in a manner that first the dephasing starts suppressing the e-ph transport channels which is maximized at $\Gamma = 0.17$ eV, and subsequently the $n_{A-LUMO}$ gets suppressed.
 
 In order to understand the resonant dephasing we have evaluated $\eta$ and $n_{phn}$ as a function of $\Gamma$ for a few e-ph coupling strength $\lambda_{e-ph}$. These are plotted in \textcolor{red}{\textit{supplementary material}}. It is found that the value of the resonant dephasing $\Gamma_{R}$ increases with the increasing $\lambda_{e-ph}$. Physically, this indicates that with increasing e-ph coupling strength one needs a faster dephasing rate to suppress the available e-ph transport channels. 
 
 (iii) For $\Gamma  \gtrsim 10^{18} \hbar$ the A-LUMO gets repopulated while the phonon modes get de-populated resulting in an increased $\eta$. This is a situation corresponding to an off-resonant dephasing where all the transport channels provided by the e-ph coupling get open again.
 
 (iv) Lastly,  at even higher dephasing rate $\Gamma  \gtrsim 10^{21} \hbar$ all the electrons are being populated at D-LUMO and D-HOMO, as can be seen from brown dot-dashed and red dashed curves corresponding to Fig. \ref{fig:occn_dephase}. This leads to a vanishing electron transport, thereby vanishing $\eta$.\\
 
\section{Summary and Conclusions}
In this manuscript, we have presented a generic model for the efficiency at maximum power of a molecular photo-cell  subjected to the influence of an environment. We consider two general forms for the molecular environment, namely (a) a localized molecular vibration coupled to the electron orbitals, and (b) local dephasing (e.g. by a continuous Zeno-like measurement of the electron orbital density), corresponding to a broad continuous spectrum of soft vibrations. We study the effect of each of these forms separately, and then the combined effect. 

The interplay between these environments and the electronic system gives rise to a rich set of effects, finger-printed in the efficiency of the molecular photo-cell. We evaluate the efficiency for a wide range of parameters, and provide detailed insight into the origins of the specific features of the efficiency.  

Our main findings are as follows: 
\begin{itemize}
    \item In the presence of dephasing, the molecular photo-cell exhibits the so-called "environment assisted quantum transport", showing a non-monotonic dependence of the efficiency on dephasing rate. The optimal dephasing rate seems to be only weakly dependent on the position of the acceptor LUMO energy level (Fig.~2). 
    \item In the absence of dephasing, the effect of a localized vibration is characterized by two regimes (Fig.~3). For weak electron-phonon coupling the system exhibits resonant behavior, the resonances correspond to coupled electron-phonon states, i.e. phonon-assisted transport. For strong electron-photon coupling, the system exhibits a line-shape similar to the classical system \cite{EDN2011,Ajisaka2015}, prompted by the high phonon occupation. 
    \item{The combination of strong electron-phonon couplings and dephasing leads to environment-hampered transport, where there is a minimum in the efficiency for some specific dephasing rate (Fig.~7).}
\end{itemize}
Our results suggest that the co-action of the classical environment and the quantum mechanical electronic system brings in distinct parameter regimes of environment assisted and environment hampered transport which, from the design perspective, implies a need for careful tuning of parameters of the real systems.  Recent advances in the experimental ability to measure the photo-voltaic conversion efficiency in single molecule junctions and PV cells make our theoretical findings experimentally verifiable \cite{Aradhya2013, Battacharyya2011, Nicholson_2010}. It would be interesting to study these competing environmental effects in natural photo-synthetic systems owing to their conceptual similarity with the HPV cells. The resonant dephasing rate corresponding to the environment hampered transport is expected to be in commensurate with the energy scales of photo-synthetic systems.

\section*{Acknowledgements}
SS would like to acknowledge the financial support provided through a Kreitman postdoctoral fellowship. This research was supported in part by the Israel Science Fund grant No.~1360/17. We are grateful to Michael Zwolak for valuable discussions.

\bibliography{heating_molecular_junctions}

\providecommand{\noopsort}[1]{}\providecommand{\singleletter}[1]{#1}
\begin{thebibliography}{43}%
\makeatletter
\providecommand \@ifxundefined [1]{%
 \@ifx{#1\undefined}
}%
\providecommand \@ifnum [1]{%
 \ifnum #1\expandafter \@firstoftwo
 \else \expandafter \@secondoftwo
 \fi
}%
\providecommand \@ifx [1]{%
 \ifx #1\expandafter \@firstoftwo
 \else \expandafter \@secondoftwo
 \fi
}%
\providecommand \natexlab [1]{#1}%
\providecommand \enquote  [1]{``#1''}%
\providecommand \bibnamefont  [1]{#1}%
\providecommand \bibfnamefont [1]{#1}%
\providecommand \citenamefont [1]{#1}%
\providecommand \href@noop [0]{\@secondoftwo}%
\providecommand \href [0]{\begingroup \@sanitize@url \@href}%
\providecommand \@href[1]{\@@startlink{#1}\@@href}%
\providecommand \@@href[1]{\endgroup#1\@@endlink}%
\providecommand \@sanitize@url [0]{\catcode `\\12\catcode `\$12\catcode
  `\&12\catcode `\#12\catcode `\^12\catcode `\_12\catcode `\%12\relax}%
\providecommand \@@startlink[1]{}%
\providecommand \@@endlink[0]{}%
\providecommand \url  [0]{\begingroup\@sanitize@url \@url }%
\providecommand \@url [1]{\endgroup\@href {#1}{\urlprefix }}%
\providecommand \urlprefix  [0]{URL }%
\providecommand \Eprint [0]{\href }%
\providecommand \doibase [0]{https://doi.org/}%
\providecommand \selectlanguage [0]{\@gobble}%
\providecommand \bibinfo  [0]{\@secondoftwo}%
\providecommand \bibfield  [0]{\@secondoftwo}%
\providecommand \translation [1]{[#1]}%
\providecommand \BibitemOpen [0]{}%
\providecommand \bibitemStop [0]{}%
\providecommand \bibitemNoStop [0]{.\EOS\space}%
\providecommand \EOS [0]{\spacefactor3000\relax}%
\providecommand \BibitemShut  [1]{\csname bibitem#1\endcsname}%
\let\auto@bib@innerbib\@empty
\bibitem [{\citenamefont {Aradhya}\ and\ \citenamefont
  {Venkataraman}(2013)}]{Aradhya2013}%
  \BibitemOpen
  \bibfield  {author} {\bibinfo {author} {\bibfnamefont {S.~V.}\ \bibnamefont
  {Aradhya}}and\ \bibinfo {author} {\bibfnamefont {L.}~\bibnamefont
  {Venkataraman}},\ }\bibfield  {title} {\bibinfo {title} {Single-molecule
  junctions beyond electronic transport},\ }\href
  {https://doi.org/10.1038/nnano.2013.91} {\bibfield  {journal} {\bibinfo
  {journal} {Nature Nanotechnology}\ }\textbf {\bibinfo {volume} {8}},\
  \bibinfo {pages} {399} (\bibinfo {year} {2013})}\BibitemShut {NoStop}%
\bibitem [{\citenamefont {Galperin}\ \emph {et~al.}(2008)\citenamefont
  {Galperin}, \citenamefont {Ratner}, \citenamefont {Nitzan},\ and\
  \citenamefont {Troisi}}]{Galperin1056}%
  \BibitemOpen
  \bibfield  {author} {\bibinfo {author} {\bibfnamefont {M.}~\bibnamefont
  {Galperin}}, \bibinfo {author} {\bibfnamefont {M.~A.}\ \bibnamefont
  {Ratner}}, \bibinfo {author} {\bibfnamefont {A.}~\bibnamefont {Nitzan}}, and\
  \bibinfo {author} {\bibfnamefont {A.}~\bibnamefont {Troisi}},\ }\bibfield
  {title} {\bibinfo {title} {Nuclear coupling and polarization in molecular
  transport junctions: Beyond tunneling to function},\ }\href
  {https://doi.org/10.1126/science.1146556} {\bibfield  {journal} {\bibinfo
  {journal} {Science}\ }\textbf {\bibinfo {volume} {319}},\ \bibinfo {pages}
  {1056} (\bibinfo {year} {2008})},\ \Eprint
  {https://arxiv.org/abs/https://science.sciencemag.org/content/319/5866/1056.full.pdf}
  {https://science.sciencemag.org/content/319/5866/1056.full.pdf} \BibitemShut
  {NoStop}%
\bibitem [{\citenamefont {Su}\ \emph {et~al.}(2016)\citenamefont {Su},
  \citenamefont {Neupane}, \citenamefont {Steigerwald}, \citenamefont
  {Venkataraman},\ and\ \citenamefont {Nuckolls}}]{Su2016}%
  \BibitemOpen
  \bibfield  {author} {\bibinfo {author} {\bibfnamefont {T.~A.}\ \bibnamefont
  {Su}}, \bibinfo {author} {\bibfnamefont {M.}~\bibnamefont {Neupane}},
  \bibinfo {author} {\bibfnamefont {M.~L.}\ \bibnamefont {Steigerwald}},
  \bibinfo {author} {\bibfnamefont {L.}~\bibnamefont {Venkataraman}}, and\
  \bibinfo {author} {\bibfnamefont {C.}~\bibnamefont {Nuckolls}},\ }\bibfield
  {title} {\bibinfo {title} {Chemical principles of single-molecule
  electronics},\ }\href {https://doi.org/10.1038/natrevmats.2016.2} {\bibfield
  {journal} {\bibinfo  {journal} {Nature Reviews Materials}\ }\textbf {\bibinfo
  {volume} {1}},\ \bibinfo {pages} {16002} (\bibinfo {year}
  {2016})}\BibitemShut {NoStop}%
\bibitem [{\citenamefont {Evers}\ and\ \citenamefont
  {Venkataraman}(2017)}]{Evers2017}%
  \BibitemOpen
  \bibfield  {author} {\bibinfo {author} {\bibfnamefont {F.}~\bibnamefont
  {Evers}}and\ \bibinfo {author} {\bibfnamefont {L.}~\bibnamefont
  {Venkataraman}},\ }\bibfield  {title} {\bibinfo {title} {Preface: Special
  topic on frontiers in molecular scale electronics},\ }\href
  {https://doi.org/10.1063/1.4977469} {\bibfield  {journal} {\bibinfo
  {journal} {The Journal of Chemical Physics}\ }\textbf {\bibinfo {volume}
  {146}},\ \bibinfo {pages} {092101} (\bibinfo {year} {2017})},\ \Eprint
  {https://arxiv.org/abs/https://doi.org/10.1063/1.4977469}
  {https://doi.org/10.1063/1.4977469} \BibitemShut {NoStop}%
\bibitem [{\citenamefont {Thoss}\ and\ \citenamefont
  {Evers}(2018)}]{Thoss2018}%
  \BibitemOpen
  \bibfield  {author} {\bibinfo {author} {\bibfnamefont {M.}~\bibnamefont
  {Thoss}}and\ \bibinfo {author} {\bibfnamefont {F.}~\bibnamefont {Evers}},\
  }\bibfield  {title} {\bibinfo {title} {Perspective: Theory of quantum
  transport in molecular junctions},\ }\href
  {https://doi.org/10.1063/1.5003306} {\bibfield  {journal} {\bibinfo
  {journal} {The Journal of Chemical Physics}\ }\textbf {\bibinfo {volume}
  {148}},\ \bibinfo {pages} {030901} (\bibinfo {year} {2018})},\ \Eprint
  {https://arxiv.org/abs/https://doi.org/10.1063/1.5003306}
  {https://doi.org/10.1063/1.5003306} \BibitemShut {NoStop}%
\bibitem [{\citenamefont {Gr{\"a}tzel}(2005)}]{Gratzel2005}%
  \BibitemOpen
  \bibfield  {author} {\bibinfo {author} {\bibfnamefont {M.}~\bibnamefont
  {Gr{\"a}tzel}},\ }\bibfield  {title} {\bibinfo {title} {Solar energy
  conversion by dye-sensitized photovoltaic cells},\ }\href
  {https://doi.org/10.1021/ic0508371} {\bibfield  {journal} {\bibinfo
  {journal} {Inorganic Chemistry}\ }\textbf {\bibinfo {volume} {44}},\ \bibinfo
  {pages} {6841} (\bibinfo {year} {2005})}\BibitemShut {NoStop}%
\bibitem [{\citenamefont {Deibel}\ and\ \citenamefont
  {Dyakonov}(2010)}]{Deibel_2010}%
  \BibitemOpen
  \bibfield  {author} {\bibinfo {author} {\bibfnamefont {C.}~\bibnamefont
  {Deibel}}and\ \bibinfo {author} {\bibfnamefont {V.}~\bibnamefont
  {Dyakonov}},\ }\bibfield  {title} {\bibinfo {title}
  {Polymer{\textendash}fullerene bulk heterojunction solar cells},\ }\href
  {https://doi.org/10.1088/0034-4885/73/9/096401} {\bibfield  {journal}
  {\bibinfo  {journal} {Reports on Progress in Physics}\ }\textbf {\bibinfo
  {volume} {73}},\ \bibinfo {pages} {096401} (\bibinfo {year}
  {2010})}\BibitemShut {NoStop}%
\bibitem [{\citenamefont {Nicholson}\ and\ \citenamefont
  {Castro}(2010)}]{Nicholson_2010}%
  \BibitemOpen
  \bibfield  {author} {\bibinfo {author} {\bibfnamefont {P.~G.}\ \bibnamefont
  {Nicholson}}and\ \bibinfo {author} {\bibfnamefont {F.~A.}\ \bibnamefont
  {Castro}},\ }\bibfield  {title} {\bibinfo {title} {Organic photovoltaics:
  principles and techniques for nanometre scale characterization},\ }\href
  {https://doi.org/10.1088/0957-4484/21/49/492001} {\bibfield  {journal}
  {\bibinfo  {journal} {Nanotechnology}\ }\textbf {\bibinfo {volume} {21}},\
  \bibinfo {pages} {492001} (\bibinfo {year} {2010})}\BibitemShut {NoStop}%
\bibitem [{\citenamefont {Einax}\ \emph {et~al.}(2011)\citenamefont {Einax},
  \citenamefont {Dierl},\ and\ \citenamefont {Nitzan}}]{EDN2011}%
  \BibitemOpen
  \bibfield  {author} {\bibinfo {author} {\bibfnamefont {M.}~\bibnamefont
  {Einax}}, \bibinfo {author} {\bibfnamefont {M.}~\bibnamefont {Dierl}}, and\
  \bibinfo {author} {\bibfnamefont {A.}~\bibnamefont {Nitzan}},\ }\bibfield
  {title} {\bibinfo {title} {Heterojunction organic photovoltaic cells as
  molecular heat engines: A simple model for the performance analysis},\ }\href
  {https://doi.org/10.1021/jp205856x} {\bibfield  {journal} {\bibinfo
  {journal} {The Journal of Physical Chemistry C}\ }\textbf {\bibinfo {volume}
  {115}},\ \bibinfo {pages} {21396} (\bibinfo {year} {2011})},\ \Eprint
  {https://arxiv.org/abs/https://doi.org/10.1021/jp205856x}
  {https://doi.org/10.1021/jp205856x} \BibitemShut {NoStop}%
\bibitem [{\citenamefont {Ajisaka}\ \emph {et~al.}(2015)\citenamefont
  {Ajisaka}, \citenamefont {Zunkovic},\ and\ \citenamefont
  {Dubi}}]{Ajisaka2015}%
  \BibitemOpen
  \bibfield  {author} {\bibinfo {author} {\bibfnamefont {S.}~\bibnamefont
  {Ajisaka}}, \bibinfo {author} {\bibfnamefont {B.}~\bibnamefont {Zunkovic}},
  and\ \bibinfo {author} {\bibfnamefont {Y.}~\bibnamefont {Dubi}},\ }\bibfield
  {title} {\bibinfo {title} {The molecular photo-cell: Quantum transport and
  energy conversion at strong non-equilibrium},\ }\href@noop {} {\bibfield
  {journal} {\bibinfo  {journal} {Scientific Reports}\ }\textbf {\bibinfo
  {volume} {5}},\ \bibinfo {pages} {8312} (\bibinfo {year} {2015})},\ \bibinfo
  {note} {article},\ \Eprint
  {https://arxiv.org/abs/https://doi.org/10.1038/srep08312}
  {https://doi.org/10.1038/srep08312} \BibitemShut {NoStop}%
\bibitem [{\citenamefont {Fruchtman}\ \emph {et~al.}(2016)\citenamefont
  {Fruchtman}, \citenamefont {G\'omez-Bombarelli}, \citenamefont {Lovett},\
  and\ \citenamefont {Gauger}}]{Fruchtman2016}%
  \BibitemOpen
  \bibfield  {author} {\bibinfo {author} {\bibfnamefont {A.}~\bibnamefont
  {Fruchtman}}, \bibinfo {author} {\bibfnamefont {R.}~\bibnamefont
  {G\'omez-Bombarelli}}, \bibinfo {author} {\bibfnamefont {B.~W.}\ \bibnamefont
  {Lovett}}, and\ \bibinfo {author} {\bibfnamefont {E.~M.}\ \bibnamefont
  {Gauger}},\ }\bibfield  {title} {\bibinfo {title} {Photocell optimization
  using dark state protection},\ }\href
  {https://doi.org/10.1103/PhysRevLett.117.203603} {\bibfield  {journal}
  {\bibinfo  {journal} {Phys. Rev. Lett.}\ }\textbf {\bibinfo {volume} {117}},\
  \bibinfo {pages} {203603} (\bibinfo {year} {2016})}\BibitemShut {NoStop}%
\bibitem [{\citenamefont {Arp}\ \emph {et~al.}(2016)\citenamefont {Arp},
  \citenamefont {Barlas}, \citenamefont {Aji},\ and\ \citenamefont
  {Gabor}}]{Arp2016}%
  \BibitemOpen
  \bibfield  {author} {\bibinfo {author} {\bibfnamefont {T.~B.}\ \bibnamefont
  {Arp}}, \bibinfo {author} {\bibfnamefont {Y.}~\bibnamefont {Barlas}},
  \bibinfo {author} {\bibfnamefont {V.}~\bibnamefont {Aji}}, and\ \bibinfo
  {author} {\bibfnamefont {N.~M.}\ \bibnamefont {Gabor}},\ }\bibfield  {title}
  {\bibinfo {title} {Natural regulation of energy flow in a green quantum
  photocell},\ }\href {https://doi.org/10.1021/acs.nanolett.6b03136} {\bibfield
   {journal} {\bibinfo  {journal} {Nano Letters}\ }\textbf {\bibinfo {volume}
  {16}},\ \bibinfo {pages} {7461} (\bibinfo {year} {2016})}\BibitemShut
  {NoStop}%
\bibitem [{\citenamefont {Killoran}\ \emph {et~al.}(2015)\citenamefont
  {Killoran}, \citenamefont {Huelga},\ and\ \citenamefont
  {Plenio}}]{Killoran2015}%
  \BibitemOpen
  \bibfield  {author} {\bibinfo {author} {\bibfnamefont {N.}~\bibnamefont
  {Killoran}}, \bibinfo {author} {\bibfnamefont {S.~F.}\ \bibnamefont
  {Huelga}}, and\ \bibinfo {author} {\bibfnamefont {M.~B.}\ \bibnamefont
  {Plenio}},\ }\bibfield  {title} {\bibinfo {title} {Enhancing light-harvesting
  power with coherent vibrational interactions: A quantum heat engine
  picture},\ }\href {https://doi.org/10.1063/1.4932307} {\bibfield  {journal}
  {\bibinfo  {journal} {The Journal of Chemical Physics}\ }\textbf {\bibinfo
  {volume} {143}},\ \bibinfo {pages} {155102} (\bibinfo {year} {2015})},\
  \Eprint {https://arxiv.org/abs/https://doi.org/10.1063/1.4932307}
  {https://doi.org/10.1063/1.4932307} \BibitemShut {NoStop}%
\bibitem [{\citenamefont {Ernzerhof}\ \emph {et~al.}(2016)\citenamefont
  {Ernzerhof}, \citenamefont {Bélanger}, \citenamefont {Mayou},\ and\
  \citenamefont {Nemati~Aram}}]{Ernzerhof2016}%
  \BibitemOpen
  \bibfield  {author} {\bibinfo {author} {\bibfnamefont {M.}~\bibnamefont
  {Ernzerhof}}, \bibinfo {author} {\bibfnamefont {M.-A.}\ \bibnamefont
  {Bélanger}}, \bibinfo {author} {\bibfnamefont {D.}~\bibnamefont {Mayou}},
  and\ \bibinfo {author} {\bibfnamefont {T.}~\bibnamefont {Nemati~Aram}},\
  }\bibfield  {title} {\bibinfo {title} {Simple model of a coherent molecular
  photocell},\ }\href {https://doi.org/10.1063/1.4944468} {\bibfield  {journal}
  {\bibinfo  {journal} {The Journal of Chemical Physics}\ }\textbf {\bibinfo
  {volume} {144}},\ \bibinfo {pages} {134102} (\bibinfo {year} {2016})},\
  \Eprint {https://arxiv.org/abs/https://doi.org/10.1063/1.4944468}
  {https://doi.org/10.1063/1.4944468} \BibitemShut {NoStop}%
\bibitem [{\citenamefont {Nemati~Aram}\ \emph {et~al.}(2016)\citenamefont
  {Nemati~Aram}, \citenamefont {Anghel-Vasilescu}, \citenamefont {Asgari},
  \citenamefont {Ernzerhof},\ and\ \citenamefont {Mayou}}]{Nemati-Aram2016}%
  \BibitemOpen
  \bibfield  {author} {\bibinfo {author} {\bibfnamefont {T.}~\bibnamefont
  {Nemati~Aram}}, \bibinfo {author} {\bibfnamefont {P.}~\bibnamefont
  {Anghel-Vasilescu}}, \bibinfo {author} {\bibfnamefont {A.}~\bibnamefont
  {Asgari}}, \bibinfo {author} {\bibfnamefont {M.}~\bibnamefont {Ernzerhof}},
  and\ \bibinfo {author} {\bibfnamefont {D.}~\bibnamefont {Mayou}},\ }\bibfield
   {title} {\bibinfo {title} {Modeling of molecular photocells: Application to
  two-level photovoltaic system with electron-hole interaction},\ }\href
  {https://doi.org/10.1063/1.4963335} {\bibfield  {journal} {\bibinfo
  {journal} {The Journal of Chemical Physics}\ }\textbf {\bibinfo {volume}
  {145}},\ \bibinfo {pages} {124116} (\bibinfo {year} {2016})},\ \Eprint
  {https://arxiv.org/abs/https://doi.org/10.1063/1.4963335}
  {https://doi.org/10.1063/1.4963335} \BibitemShut {NoStop}%
\bibitem [{\citenamefont {Nemati~Aram}\ \emph {et~al.}(2017)\citenamefont
  {Nemati~Aram}, \citenamefont {Ernzerhof}, \citenamefont {Asgari},\ and\
  \citenamefont {Mayou}}]{Nemati-Aram2017}%
  \BibitemOpen
  \bibfield  {author} {\bibinfo {author} {\bibfnamefont {T.}~\bibnamefont
  {Nemati~Aram}}, \bibinfo {author} {\bibfnamefont {M.}~\bibnamefont
  {Ernzerhof}}, \bibinfo {author} {\bibfnamefont {A.}~\bibnamefont {Asgari}},
  and\ \bibinfo {author} {\bibfnamefont {D.}~\bibnamefont {Mayou}},\ }\bibfield
   {title} {\bibinfo {title} {The impact of long-range electron-hole
  interaction on the charge separation yield of molecular photocells},\ }\href
  {https://doi.org/10.1063/1.4973984} {\bibfield  {journal} {\bibinfo
  {journal} {The Journal of Chemical Physics}\ }\textbf {\bibinfo {volume}
  {146}},\ \bibinfo {pages} {034103} (\bibinfo {year} {2017})},\ \Eprint
  {https://arxiv.org/abs/https://doi.org/10.1063/1.4973984}
  {https://doi.org/10.1063/1.4973984} \BibitemShut {NoStop}%
\bibitem [{\citenamefont {{Aram, Tahereh Nemati}}\ \emph
  {et~al.}(2017)\citenamefont {{Aram, Tahereh Nemati}}, \citenamefont {{Asgari,
  Asghar}}, \citenamefont {{Ernzerhof, Matthias}}, \citenamefont {{Qu\'emerais,
  Pascal}},\ and\ \citenamefont {{Mayou, Didier}}}]{Aram2017}%
  \BibitemOpen
  \bibfield  {author} {\bibinfo {author} {\bibnamefont {{Aram, Tahereh
  Nemati}}}, \bibinfo {author} {\bibnamefont {{Asgari, Asghar}}}, \bibinfo
  {author} {\bibnamefont {{Ernzerhof, Matthias}}}, \bibinfo {author}
  {\bibnamefont {{Qu\'emerais, Pascal}}}, and\ \bibinfo {author} {\bibnamefont
  {{Mayou, Didier}}},\ }\bibfield  {title} {\bibinfo {title} {Quantum modeling
  of two-level photovoltaic systems},\ }\href
  {https://doi.org/10.1051/epjpv/2017004} {\bibfield  {journal} {\bibinfo
  {journal} {EPJ Photovolt.}\ }\textbf {\bibinfo {volume} {8}},\ \bibinfo
  {pages} {85503} (\bibinfo {year} {2017})}\BibitemShut {NoStop}%
\bibitem [{\citenamefont {Ho}(2002)}]{Ho2002}%
  \BibitemOpen
  \bibfield  {author} {\bibinfo {author} {\bibfnamefont {W.}~\bibnamefont
  {Ho}},\ }\bibfield  {title} {\bibinfo {title} {Single-molecule chemistry},\
  }\href {https://doi.org/10.1063/1.1521153} {\bibfield  {journal} {\bibinfo
  {journal} {The Journal of Chemical Physics}\ }\textbf {\bibinfo {volume}
  {117}},\ \bibinfo {pages} {11033} (\bibinfo {year} {2002})},\ \Eprint
  {https://arxiv.org/abs/https://doi.org/10.1063/1.1521153}
  {https://doi.org/10.1063/1.1521153} \BibitemShut {NoStop}%
\bibitem [{\citenamefont {Nitzan}(2001)}]{NitzanReview2001}%
  \BibitemOpen
  \bibfield  {author} {\bibinfo {author} {\bibfnamefont {A.}~\bibnamefont
  {Nitzan}},\ }\bibfield  {title} {\bibinfo {title} {Electron transmission
  through molecules and molecular interfaces},\ }\href
  {https://doi.org/10.1146/annurev.physchem.52.1.681} {\bibfield  {journal}
  {\bibinfo  {journal} {Annual Review of Physical Chemistry}\ }\textbf
  {\bibinfo {volume} {52}},\ \bibinfo {pages} {681} (\bibinfo {year} {2001})},\
  \bibinfo {note} {pMID: 11326078},\ \Eprint
  {https://arxiv.org/abs/https://doi.org/10.1146/annurev.physchem.52.1.681}
  {https://doi.org/10.1146/annurev.physchem.52.1.681} \BibitemShut {NoStop}%
\bibitem [{\citenamefont {Galperin}\ \emph {et~al.}(2007)\citenamefont
  {Galperin}, \citenamefont {Ratner},\ and\ \citenamefont
  {Nitzan}}]{Galperin_2007}%
  \BibitemOpen
  \bibfield  {author} {\bibinfo {author} {\bibfnamefont {M.}~\bibnamefont
  {Galperin}}, \bibinfo {author} {\bibfnamefont {M.~A.}\ \bibnamefont
  {Ratner}}, and\ \bibinfo {author} {\bibfnamefont {A.}~\bibnamefont
  {Nitzan}},\ }\bibfield  {title} {\bibinfo {title} {Molecular transport
  junctions: vibrational effects},\ }\href
  {https://doi.org/10.1088/0953-8984/19/10/103201} {\bibfield  {journal}
  {\bibinfo  {journal} {Journal of Physics: Condensed Matter}\ }\textbf
  {\bibinfo {volume} {19}},\ \bibinfo {pages} {103201} (\bibinfo {year}
  {2007})}\BibitemShut {NoStop}%
\bibitem [{\citenamefont {Qin}\ \emph {et~al.}(2017)\citenamefont {Qin},
  \citenamefont {Shen}, \citenamefont {Zhao},\ and\ \citenamefont
  {Yi}}]{Qin_Shen_Zhao_Yi_PhysRevA}%
  \BibitemOpen
  \bibfield  {author} {\bibinfo {author} {\bibfnamefont {M.}~\bibnamefont
  {Qin}}, \bibinfo {author} {\bibfnamefont {H.~Z.}\ \bibnamefont {Shen}},
  \bibinfo {author} {\bibfnamefont {X.~L.}\ \bibnamefont {Zhao}}, and\ \bibinfo
  {author} {\bibfnamefont {X.~X.}\ \bibnamefont {Yi}},\ }\bibfield  {title}
  {\bibinfo {title} {Effects of system-bath coupling on a photosynthetic heat
  engine: A polaron master-equation approach},\ }\href
  {https://doi.org/10.1103/PhysRevA.96.012125} {\bibfield  {journal} {\bibinfo
  {journal} {Phys. Rev. A}\ }\textbf {\bibinfo {volume} {96}},\ \bibinfo
  {pages} {012125} (\bibinfo {year} {2017})}\BibitemShut {NoStop}%
\bibitem [{\citenamefont {Qin}\ \emph {et~al.}(2019)\citenamefont {Qin},
  \citenamefont {Wang}, \citenamefont {Cui},\ and\ \citenamefont
  {Yi}}]{Qin_Wang_Cui_Yi_PhysRevA}%
  \BibitemOpen
  \bibfield  {author} {\bibinfo {author} {\bibfnamefont {M.}~\bibnamefont
  {Qin}}, \bibinfo {author} {\bibfnamefont {C.~Y.}\ \bibnamefont {Wang}},
  \bibinfo {author} {\bibfnamefont {H.~T.}\ \bibnamefont {Cui}}, and\ \bibinfo
  {author} {\bibfnamefont {X.~X.}\ \bibnamefont {Yi}},\ }\bibfield  {title}
  {\bibinfo {title} {Excitation-energy transfer from weak to strong coupling:
  Role of initial system-bath correlations},\ }\href
  {https://doi.org/10.1103/PhysRevA.99.032111} {\bibfield  {journal} {\bibinfo
  {journal} {Phys. Rev. A}\ }\textbf {\bibinfo {volume} {99}},\ \bibinfo
  {pages} {032111} (\bibinfo {year} {2019})}\BibitemShut {NoStop}%
\bibitem [{\citenamefont {Qin}\ \emph {et~al.}(2016)\citenamefont {Qin},
  \citenamefont {Shen},\ and\ \citenamefont {Yi}}]{Qin_Shen_Yi_JChemPhys}%
  \BibitemOpen
  \bibfield  {author} {\bibinfo {author} {\bibfnamefont {M.}~\bibnamefont
  {Qin}}, \bibinfo {author} {\bibfnamefont {H.~Z.}\ \bibnamefont {Shen}}, and\
  \bibinfo {author} {\bibfnamefont {X.~X.}\ \bibnamefont {Yi}},\ }\bibfield
  {title} {\bibinfo {title} {A multi-pathway model for photosynthetic reaction
  center},\ }\href {https://doi.org/10.1063/1.4944730} {\bibfield  {journal}
  {\bibinfo  {journal} {The Journal of Chemical Physics}\ }\textbf {\bibinfo
  {volume} {144}},\ \bibinfo {pages} {125103} (\bibinfo {year} {2016})},\
  \Eprint {https://arxiv.org/abs/https://doi.org/10.1063/1.4944730}
  {https://doi.org/10.1063/1.4944730} \BibitemShut {NoStop}%
\bibitem [{\citenamefont {Ballmann}\ \emph {et~al.}(2012)\citenamefont
  {Ballmann}, \citenamefont {H\"artle}, \citenamefont {Coto}, \citenamefont
  {Elbing}, \citenamefont {Mayor}, \citenamefont {Bryce}, \citenamefont
  {Thoss},\ and\ \citenamefont {Weber}}]{Ballmann2012}%
  \BibitemOpen
  \bibfield  {author} {\bibinfo {author} {\bibfnamefont {S.}~\bibnamefont
  {Ballmann}}, \bibinfo {author} {\bibfnamefont {R.}~\bibnamefont {H\"artle}},
  \bibinfo {author} {\bibfnamefont {P.~B.}\ \bibnamefont {Coto}}, \bibinfo
  {author} {\bibfnamefont {M.}~\bibnamefont {Elbing}}, \bibinfo {author}
  {\bibfnamefont {M.}~\bibnamefont {Mayor}}, \bibinfo {author} {\bibfnamefont
  {M.~R.}\ \bibnamefont {Bryce}}, \bibinfo {author} {\bibfnamefont
  {M.}~\bibnamefont {Thoss}}, and\ \bibinfo {author} {\bibfnamefont {H.~B.}\
  \bibnamefont {Weber}},\ }\bibfield  {title} {\bibinfo {title} {Experimental
  evidence for quantum interference and vibrationally induced decoherence in
  single-molecule junctions},\ }\href
  {https://doi.org/10.1103/PhysRevLett.109.056801} {\bibfield  {journal}
  {\bibinfo  {journal} {Phys. Rev. Lett.}\ }\textbf {\bibinfo {volume} {109}},\
  \bibinfo {pages} {056801} (\bibinfo {year} {2012})}\BibitemShut {NoStop}%
\bibitem [{\citenamefont {H\"artle}\ \emph {et~al.}(2011)\citenamefont
  {H\"artle}, \citenamefont {Butzin}, \citenamefont {Rubio-Pons},\ and\
  \citenamefont {Thoss}}]{Hartle2011}%
  \BibitemOpen
  \bibfield  {author} {\bibinfo {author} {\bibfnamefont {R.}~\bibnamefont
  {H\"artle}}, \bibinfo {author} {\bibfnamefont {M.}~\bibnamefont {Butzin}},
  \bibinfo {author} {\bibfnamefont {O.}~\bibnamefont {Rubio-Pons}}, and\
  \bibinfo {author} {\bibfnamefont {M.}~\bibnamefont {Thoss}},\ }\bibfield
  {title} {\bibinfo {title} {Quantum interference and decoherence in
  single-molecule junctions: How vibrations induce electrical current},\ }\href
  {https://doi.org/10.1103/PhysRevLett.107.046802} {\bibfield  {journal}
  {\bibinfo  {journal} {Phys. Rev. Lett.}\ }\textbf {\bibinfo {volume} {107}},\
  \bibinfo {pages} {046802} (\bibinfo {year} {2011})}\BibitemShut {NoStop}%
\bibitem [{\citenamefont {Sowa}\ \emph {et~al.}(2017)\citenamefont {Sowa},
  \citenamefont {Mol}, \citenamefont {Briggs},\ and\ \citenamefont
  {Gauger}}]{SowaRSC2017}%
  \BibitemOpen
  \bibfield  {author} {\bibinfo {author} {\bibfnamefont {J.~K.}\ \bibnamefont
  {Sowa}}, \bibinfo {author} {\bibfnamefont {J.~A.}\ \bibnamefont {Mol}},
  \bibinfo {author} {\bibfnamefont {G.~A.~D.}\ \bibnamefont {Briggs}}, and\
  \bibinfo {author} {\bibfnamefont {E.~M.}\ \bibnamefont {Gauger}},\ }\bibfield
   {title} {\bibinfo {title} {Environment-assisted quantum transport through
  single-molecule junctions},\ }\href {https://doi.org/10.1039/C7CP06237K}
  {\bibfield  {journal} {\bibinfo  {journal} {Phys. Chem. Chem. Phys.}\
  }\textbf {\bibinfo {volume} {19}},\ \bibinfo {pages} {29534} (\bibinfo {year}
  {2017})}\BibitemShut {NoStop}%
\bibitem [{\citenamefont {Plenio}\ and\ \citenamefont
  {Huelga}(2008)}]{Plenio_2008}%
  \BibitemOpen
  \bibfield  {author} {\bibinfo {author} {\bibfnamefont {M.~B.}\ \bibnamefont
  {Plenio}}and\ \bibinfo {author} {\bibfnamefont {S.~F.}\ \bibnamefont
  {Huelga}},\ }\bibfield  {title} {\bibinfo {title} {Dephasing-assisted
  transport: quantum networks and biomolecules},\ }\href
  {https://doi.org/10.1088/1367-2630/10/11/113019} {\bibfield  {journal}
  {\bibinfo  {journal} {New Journal of Physics}\ }\textbf {\bibinfo {volume}
  {10}},\ \bibinfo {pages} {113019} (\bibinfo {year} {2008})}\BibitemShut
  {NoStop}%
\bibitem [{\citenamefont {Mohseni}\ \emph {et~al.}(2008)\citenamefont
  {Mohseni}, \citenamefont {Rebentrost}, \citenamefont {Lloyd},\ and\
  \citenamefont {Aspuru-Guzik}}]{Mohseni2008}%
  \BibitemOpen
  \bibfield  {author} {\bibinfo {author} {\bibfnamefont {M.}~\bibnamefont
  {Mohseni}}, \bibinfo {author} {\bibfnamefont {P.}~\bibnamefont {Rebentrost}},
  \bibinfo {author} {\bibfnamefont {S.}~\bibnamefont {Lloyd}}, and\ \bibinfo
  {author} {\bibfnamefont {A.}~\bibnamefont {Aspuru-Guzik}},\ }\bibfield
  {title} {\bibinfo {title} {Environment-assisted quantum walks in
  photosynthetic energy transfer},\ }\href {https://doi.org/10.1063/1.3002335}
  {\bibfield  {journal} {\bibinfo  {journal} {The Journal of Chemical Physics}\
  }\textbf {\bibinfo {volume} {129}},\ \bibinfo {pages} {174106} (\bibinfo
  {year} {2008})},\ \Eprint
  {https://arxiv.org/abs/https://doi.org/10.1063/1.3002335}
  {https://doi.org/10.1063/1.3002335} \BibitemShut {NoStop}%
\bibitem [{\citenamefont {Rebentrost}\ \emph {et~al.}(2009)\citenamefont
  {Rebentrost}, \citenamefont {Mohseni}, \citenamefont {Kassal}, \citenamefont
  {Lloyd},\ and\ \citenamefont {Aspuru-Guzik}}]{Rebentrost_2009}%
  \BibitemOpen
  \bibfield  {author} {\bibinfo {author} {\bibfnamefont {P.}~\bibnamefont
  {Rebentrost}}, \bibinfo {author} {\bibfnamefont {M.}~\bibnamefont {Mohseni}},
  \bibinfo {author} {\bibfnamefont {I.}~\bibnamefont {Kassal}}, \bibinfo
  {author} {\bibfnamefont {S.}~\bibnamefont {Lloyd}}, and\ \bibinfo {author}
  {\bibfnamefont {A.}~\bibnamefont {Aspuru-Guzik}},\ }\bibfield  {title}
  {\bibinfo {title} {Environment-assisted quantum transport},\ }\href
  {https://doi.org/10.1088/1367-2630/11/3/033003} {\bibfield  {journal}
  {\bibinfo  {journal} {New Journal of Physics}\ }\textbf {\bibinfo {volume}
  {11}},\ \bibinfo {pages} {033003} (\bibinfo {year} {2009})}\BibitemShut
  {NoStop}%
\bibitem [{\citenamefont {Caruso}\ \emph {et~al.}(2009)\citenamefont {Caruso},
  \citenamefont {Chin}, \citenamefont {Datta}, \citenamefont {Huelga},\ and\
  \citenamefont {Plenio}}]{Caruso_J_Chem_phys}%
  \BibitemOpen
  \bibfield  {author} {\bibinfo {author} {\bibfnamefont {F.}~\bibnamefont
  {Caruso}}, \bibinfo {author} {\bibfnamefont {A.~W.}\ \bibnamefont {Chin}},
  \bibinfo {author} {\bibfnamefont {A.}~\bibnamefont {Datta}}, \bibinfo
  {author} {\bibfnamefont {S.~F.}\ \bibnamefont {Huelga}}, and\ \bibinfo
  {author} {\bibfnamefont {M.~B.}\ \bibnamefont {Plenio}},\ }\bibfield  {title}
  {\bibinfo {title} {Highly efficient energy excitation transfer in
  light-harvesting complexes: The fundamental role of noise-assisted
  transport},\ }\href {https://doi.org/10.1063/1.3223548} {\bibfield  {journal}
  {\bibinfo  {journal} {The Journal of Chemical Physics}\ }\textbf {\bibinfo
  {volume} {131}},\ \bibinfo {pages} {105106} (\bibinfo {year} {2009})},\
  \Eprint
  {https://arxiv.org/abs/https://aip.scitation.org/doi/pdf/10.1063/1.3223548}
  {https://aip.scitation.org/doi/pdf/10.1063/1.3223548} \BibitemShut {NoStop}%
\bibitem [{\citenamefont {Zerah-Harush}\ and\ \citenamefont
  {Dubi}(2018)}]{Zerah-Harush2018}%
  \BibitemOpen
  \bibfield  {author} {\bibinfo {author} {\bibfnamefont {E.}~\bibnamefont
  {Zerah-Harush}}and\ \bibinfo {author} {\bibfnamefont {Y.}~\bibnamefont
  {Dubi}},\ }\bibfield  {title} {\bibinfo {title} {Universal origin for
  environment-assisted quantum transport in exciton transfer networks},\ }\href
  {https://doi.org/10.1021/acs.jpclett.7b03306} {\bibfield  {journal} {\bibinfo
   {journal} {The Journal of Physical Chemistry Letters}\ }\textbf {\bibinfo
  {volume} {9}},\ \bibinfo {pages} {1689} (\bibinfo {year} {2018})}\BibitemShut
  {NoStop}%
\bibitem [{\citenamefont {Maier}\ \emph {et~al.}(2019)\citenamefont {Maier},
  \citenamefont {Brydges}, \citenamefont {Jurcevic}, \citenamefont {Trautmann},
  \citenamefont {Hempel}, \citenamefont {Lanyon}, \citenamefont {Hauke},
  \citenamefont {Blatt},\ and\ \citenamefont {Roos}}]{Maier2019}%
  \BibitemOpen
  \bibfield  {author} {\bibinfo {author} {\bibfnamefont {C.}~\bibnamefont
  {Maier}}, \bibinfo {author} {\bibfnamefont {T.}~\bibnamefont {Brydges}},
  \bibinfo {author} {\bibfnamefont {P.}~\bibnamefont {Jurcevic}}, \bibinfo
  {author} {\bibfnamefont {N.}~\bibnamefont {Trautmann}}, \bibinfo {author}
  {\bibfnamefont {C.}~\bibnamefont {Hempel}}, \bibinfo {author} {\bibfnamefont
  {B.~P.}\ \bibnamefont {Lanyon}}, \bibinfo {author} {\bibfnamefont
  {P.}~\bibnamefont {Hauke}}, \bibinfo {author} {\bibfnamefont
  {R.}~\bibnamefont {Blatt}}, and\ \bibinfo {author} {\bibfnamefont {C.~F.}\
  \bibnamefont {Roos}},\ }\bibfield  {title} {\bibinfo {title}
  {Environment-assisted quantum transport in a 10-qubit network},\ }\href
  {https://doi.org/10.1103/PhysRevLett.122.050501} {\bibfield  {journal}
  {\bibinfo  {journal} {Phys. Rev. Lett.}\ }\textbf {\bibinfo {volume} {122}},\
  \bibinfo {pages} {050501} (\bibinfo {year} {2019})}\BibitemShut {NoStop}%
\bibitem [{\citenamefont {Gorman}\ \emph {et~al.}(2018)\citenamefont {Gorman},
  \citenamefont {Hemmerling}, \citenamefont {Megidish}, \citenamefont
  {Moeller}, \citenamefont {Schindler}, \citenamefont {Sarovar},\ and\
  \citenamefont {Haeffner}}]{Gorman2018}%
  \BibitemOpen
  \bibfield  {author} {\bibinfo {author} {\bibfnamefont {D.~J.}\ \bibnamefont
  {Gorman}}, \bibinfo {author} {\bibfnamefont {B.}~\bibnamefont {Hemmerling}},
  \bibinfo {author} {\bibfnamefont {E.}~\bibnamefont {Megidish}}, \bibinfo
  {author} {\bibfnamefont {S.~A.}\ \bibnamefont {Moeller}}, \bibinfo {author}
  {\bibfnamefont {P.}~\bibnamefont {Schindler}}, \bibinfo {author}
  {\bibfnamefont {M.}~\bibnamefont {Sarovar}}, and\ \bibinfo {author}
  {\bibfnamefont {H.}~\bibnamefont {Haeffner}},\ }\bibfield  {title} {\bibinfo
  {title} {Engineering vibrationally assisted energy transfer in a trapped-ion
  quantum simulator},\ }\href {https://doi.org/10.1103/PhysRevX.8.011038}
  {\bibfield  {journal} {\bibinfo  {journal} {Phys. Rev. X}\ }\textbf {\bibinfo
  {volume} {8}},\ \bibinfo {pages} {011038} (\bibinfo {year}
  {2018})}\BibitemShut {NoStop}%
\bibitem [{\citenamefont {Viciani}\ \emph {et~al.}(2015)\citenamefont
  {Viciani}, \citenamefont {Lima}, \citenamefont {Bellini},\ and\ \citenamefont
  {Caruso}}]{Viciani2015}%
  \BibitemOpen
  \bibfield  {author} {\bibinfo {author} {\bibfnamefont {S.}~\bibnamefont
  {Viciani}}, \bibinfo {author} {\bibfnamefont {M.}~\bibnamefont {Lima}},
  \bibinfo {author} {\bibfnamefont {M.}~\bibnamefont {Bellini}}, and\ \bibinfo
  {author} {\bibfnamefont {F.}~\bibnamefont {Caruso}},\ }\bibfield  {title}
  {\bibinfo {title} {Observation of noise-assisted transport in an all-optical
  cavity-based network},\ }\href
  {https://doi.org/10.1103/PhysRevLett.115.083601} {\bibfield  {journal}
  {\bibinfo  {journal} {Phys. Rev. Lett.}\ }\textbf {\bibinfo {volume} {115}},\
  \bibinfo {pages} {083601} (\bibinfo {year} {2015})}\BibitemShut {NoStop}%
\bibitem [{\citenamefont {Lindblad}(1976)}]{Lindblad}%
  \BibitemOpen
  \bibfield  {author} {\bibinfo {author} {\bibfnamefont {G.}~\bibnamefont
  {Lindblad}},\ }\bibfield  {title} {\bibinfo {title} {On the generators of
  quantum dynamical semigroups},\ }\href {https://doi.org/10.1007/BF01608499}
  {\bibfield  {journal} {\bibinfo  {journal} {Commun.Math. Phys.}\ }\textbf
  {\bibinfo {volume} {48}},\ \bibinfo {pages} {119} (\bibinfo {year}
  {(1976)})}\BibitemShut {NoStop}%
\bibitem [{\citenamefont {Gorini}\ \emph {et~al.}(1976)\citenamefont {Gorini},
  \citenamefont {Kossakowski},\ and\ \citenamefont {Sudarshan}}]{GKS}%
  \BibitemOpen
  \bibfield  {author} {\bibinfo {author} {\bibfnamefont {V.}~\bibnamefont
  {Gorini}}, \bibinfo {author} {\bibfnamefont {A.}~\bibnamefont {Kossakowski}},
  and\ \bibinfo {author} {\bibfnamefont {E.~C.~G.}\ \bibnamefont {Sudarshan}},\
  }\bibfield  {title} {\bibinfo {title} {Completely positive dynamical
  semigroups of n‐level systems},\ }\href {https://doi.org/10.1063/1.522979}
  {\bibfield  {journal} {\bibinfo  {journal} {Journal of Mathematical Physics}\
  }\textbf {\bibinfo {volume} {17}},\ \bibinfo {pages} {821} (\bibinfo {year}
  {1976})},\ \Eprint
  {https://arxiv.org/abs/https://aip.scitation.org/doi/pdf/10.1063/1.522979}
  {https://aip.scitation.org/doi/pdf/10.1063/1.522979} \BibitemShut {NoStop}%
\bibitem [{\citenamefont {Breuer}\ and\ \citenamefont
  {Petruccione}(2004)}]{Breuer}%
  \BibitemOpen
  \bibfield  {author} {\bibinfo {author} {\bibfnamefont {H.-P.}\ \bibnamefont
  {Breuer}}and\ \bibinfo {author} {\bibfnamefont {F.}~\bibnamefont
  {Petruccione}},\ }\bibfield  {title} {\bibinfo {title} {Quantum master
  equations},\ }in\ \href@noop {} {\emph {\bibinfo {booktitle} {The Theory of
  Open Quantum Systems}}}\ (\bibinfo  {publisher} {Oxford University Press},\
  \bibinfo {address} {Oxford},\ \bibinfo {year} {2004})\ Chap.~\bibinfo
  {chapter} {3}, p.\ \bibinfo {pages} {117}\BibitemShut {NoStop}%
\bibitem [{\citenamefont {Purkayastha}\ \emph {et~al.}(2016)\citenamefont
  {Purkayastha}, \citenamefont {Dhar},\ and\ \citenamefont
  {Kulkarni}}]{Archak_Dhar_Kulkarni_PhysRevA}%
  \BibitemOpen
  \bibfield  {author} {\bibinfo {author} {\bibfnamefont {A.}~\bibnamefont
  {Purkayastha}}, \bibinfo {author} {\bibfnamefont {A.}~\bibnamefont {Dhar}},
  and\ \bibinfo {author} {\bibfnamefont {M.}~\bibnamefont {Kulkarni}},\
  }\bibfield  {title} {\bibinfo {title} {Out-of-equilibrium open quantum
  systems: A comparison of approximate quantum master equation approaches with
  exact results},\ }\href {https://doi.org/10.1103/PhysRevA.93.062114}
  {\bibfield  {journal} {\bibinfo  {journal} {Phys. Rev. A}\ }\textbf {\bibinfo
  {volume} {93}},\ \bibinfo {pages} {062114} (\bibinfo {year}
  {2016})}\BibitemShut {NoStop}%
\bibitem [{\citenamefont {Esposito}\ \emph {et~al.}(2009)\citenamefont
  {Esposito}, \citenamefont {Lindenberg},\ and\ \citenamefont {Van~den
  Broeck}}]{EspositoPRL}%
  \BibitemOpen
  \bibfield  {author} {\bibinfo {author} {\bibfnamefont {M.}~\bibnamefont
  {Esposito}}, \bibinfo {author} {\bibfnamefont {K.}~\bibnamefont
  {Lindenberg}}, and\ \bibinfo {author} {\bibfnamefont {C.}~\bibnamefont
  {Van~den Broeck}},\ }\bibfield  {title} {\bibinfo {title} {Universality of
  efficiency at maximum power},\ }\href
  {https://doi.org/10.1103/PhysRevLett.102.130602} {\bibfield  {journal}
  {\bibinfo  {journal} {Phys. Rev. Lett.}\ }\textbf {\bibinfo {volume} {102}},\
  \bibinfo {pages} {130602} (\bibinfo {year} {2009})}\BibitemShut {NoStop}%
\bibitem [{\citenamefont {Van~den Broeck}(2005)}]{VandenBroeckPRL}%
  \BibitemOpen
  \bibfield  {author} {\bibinfo {author} {\bibfnamefont {C.}~\bibnamefont
  {Van~den Broeck}},\ }\bibfield  {title} {\bibinfo {title} {Thermodynamic
  efficiency at maximum power},\ }\href
  {https://doi.org/10.1103/PhysRevLett.95.190602} {\bibfield  {journal}
  {\bibinfo  {journal} {Phys. Rev. Lett.}\ }\textbf {\bibinfo {volume} {95}},\
  \bibinfo {pages} {190602} (\bibinfo {year} {2005})}\BibitemShut {NoStop}%
\bibitem [{\citenamefont {Dziarmaga}\ \emph {et~al.}(2012)\citenamefont
  {Dziarmaga}, \citenamefont {Zurek},\ and\ \citenamefont
  {Zwolak}}]{dziarmaga2012non}%
  \BibitemOpen
  \bibfield  {author} {\bibinfo {author} {\bibfnamefont {J.}~\bibnamefont
  {Dziarmaga}}, \bibinfo {author} {\bibfnamefont {W.~H.}\ \bibnamefont
  {Zurek}}, and\ \bibinfo {author} {\bibfnamefont {M.}~\bibnamefont {Zwolak}},\
  }\bibfield  {title} {\bibinfo {title} {Non-local quantum superpositions of
  topological defects},\ }\href@noop {} {\bibfield  {journal} {\bibinfo
  {journal} {Nature Physics}\ }\textbf {\bibinfo {volume} {8}},\ \bibinfo
  {pages} {49} (\bibinfo {year} {2012})}\BibitemShut {NoStop}%
\bibitem [{\citenamefont {Zerah-Harush}\ and\ \citenamefont
  {Dubi}(2019)}]{zerahharush2019environmentassisted}%
  \BibitemOpen
  \bibfield  {author} {\bibinfo {author} {\bibfnamefont {E.}~\bibnamefont
  {Zerah-Harush}}and\ \bibinfo {author} {\bibfnamefont {Y.}~\bibnamefont
  {Dubi}},\ }\href@noop {} {\bibinfo {title} {Environment-assisted quantum
  photo-protection in interacting multi-exciton transfer complexes}} (\bibinfo
  {year} {2019}),\ \Eprint {https://arxiv.org/abs/1906.10231} {arXiv:1906.10231
  [physics.bio-ph]} \BibitemShut {NoStop}%
\bibitem [{\citenamefont {Battacharyya}\ \emph {et~al.}(2011)\citenamefont
  {Battacharyya}, \citenamefont {Kibel}, \citenamefont {Kodis}, \citenamefont
  {Liddell}, \citenamefont {Gervaldo}, \citenamefont {Gust},\ and\
  \citenamefont {Lindsay}}]{Battacharyya2011}%
  \BibitemOpen
  \bibfield  {author} {\bibinfo {author} {\bibfnamefont {S.}~\bibnamefont
  {Battacharyya}}, \bibinfo {author} {\bibfnamefont {A.}~\bibnamefont {Kibel}},
  \bibinfo {author} {\bibfnamefont {G.}~\bibnamefont {Kodis}}, \bibinfo
  {author} {\bibfnamefont {P.~A.}\ \bibnamefont {Liddell}}, \bibinfo {author}
  {\bibfnamefont {M.}~\bibnamefont {Gervaldo}}, \bibinfo {author}
  {\bibfnamefont {D.}~\bibnamefont {Gust}}, and\ \bibinfo {author}
  {\bibfnamefont {S.}~\bibnamefont {Lindsay}},\ }\bibfield  {title} {\bibinfo
  {title} {Optical modulation of molecular conductance},\ }\href
  {https://doi.org/10.1021/nl200977c} {\bibfield  {journal} {\bibinfo
  {journal} {Nano Letters}\ }\textbf {\bibinfo {volume} {11}},\ \bibinfo
  {pages} {2709} (\bibinfo {year} {2011})}\BibitemShut {NoStop}%
\end{thebibliography}%

\end{document}